\documentclass[a4paper,11pt]{article}
\usepackage{graphicx,amssymb,bm,latexsym,color,epsf}
\makeatletter
\@addtoreset{equation}{section}

\makeatother
\newcommand{\be}{\begin{equation}}
\newcommand{\ee}{\end{equation}}
\newcommand{\bea}{\begin{eqnarray}}
\newcommand{\ena}{\end{eqnarray}}
\newcommand{\vs}[1]{\vspace{#1 mm}}
\newcommand{\hs}[1]{\hspace{#1 mm}}
\renewcommand{\a}{\alpha}
\renewcommand{\b}{\beta}
\renewcommand{\c}{\gamma}
\newcommand{\G}{\Gamma}
\renewcommand{\d}{\delta}

\newcommand{\s}{\sigma}
\renewcommand{\t}{\theta}
\newcommand{\vp}{\varphi}
\newcommand{\la}{\lambda}

\newcommand{\nn}{\nonumber\\}
\newcommand{\p}[1]{(\ref{#1})}

\newcommand{\br}{\bar R}
\newcommand{\bR}{\bar R}
\newcommand{\bg}{\bar g}

\newcommand{\bnabla}{\bar\nabla}

\newcommand{\Tr}{{\rm Tr}}

\def\eq#1{(\ref{#1})}

\begin{document}

\begin{titlepage}

\renewcommand{\thefootnote}{\fnsymbol{footnote}}
\begin{flushright}
KU-TP 067 \\
\today
\end{flushright}

\vs{10}
\begin{center}
{\Large\bf Renormalization Group Equation for $f(R)$ gravity \\
 on hyperbolic spaces}
\vs{15}

{\large
Kevin Falls$^{a,}$\footnote{e-mail address: k.falls@thphys.uni-heidelberg.de}
and
Nobuyoshi Ohta$^{b,}$\footnote{e-mail address: ohtan@phys.kindai.ac.jp}
} \\
\vs{10}

$^a${\em Instit\"{u}t f\"{u}r Theoretische Physik, Universit\"{a}t Heidelberg, Philosophenweg 12,
 69120 Heidelberg, Germany}
\\ \vs{2}
$^b${\em Department of Physics, Kindai University,
Higashi-Osaka, Osaka 577-8502, Japan}

\vs{15}
{\bf Abstract}
\end{center}

We derive the flow equation for the gravitational effective average action
in an $f(R)$ truncation on hyperbolic spacetimes using the exponential parametrization of the metric.
In contrast to previous works on compact spaces, we are able to evaluate traces exactly using
the optimised cutoff.
This reveals in particular that all modes can be integrated out for a finite value of the cutoff
due to a gap in the spectrum of the Laplacian, leading to the effective action.
Studying polynomial solutions, we find poorer convergence than has been found on compact spacetimes
even though at small curvature the equations only differ in the treatment of certain modes.
In the vicinity of an asymptotically free fixed point, we find the universal beta function
for the $R^2$ coupling and compute the corresponding effective action which involves
an $R^2 \log (R^2)$ quantum correction.


\end{titlepage}
\newpage
\setcounter{page}{2}
\renewcommand{\thefootnote}{\arabic{footnote}}
\setcounter{footnote}{0}

\section{Introduction}

There are various approaches to the formulation of quantum gravity.
Whether one considers Einstein gravity or some other formulation, such as string theory, 
it is known that quantum effects generically introduce higher order terms
in the curvature. In such cases, it is quite often assumed that higher order terms are just
perturbative corrections and should not play any dominant role in the quantum theories.
However, it is natural that the higher order terms are more important than the Einstein term
in the high energy region as long as they are present. Indeed it has long been known that Einstein
gravity theory with quadratic curvature terms is renormalizable due precisely to the higher order
terms~\cite{Stelle}. On the other hand it has also been known that this class of theories suffer from
the problem of ghosts and the important property of the unitarity is not maintained.
For this reason, these theories were abandoned for some time.

However these problems may be circumvented if we go beyond a purely perturbative
approach. The asymptotic safety program is one of the promising candidates within
the conventional field theory viewpoints, where an interacting ultraviolet (UV) complete theory
is searched for as a fixed point (FP) of the functional renormalization group equation (FRGE)
in theory space~\cite{W}.
If we can find a nontrivial fixed point at high energies, this would define the quantum
effective action without divergences in terms of the FRGE and therefore provide a continuum limit
for quantum gravity. In fact a particular realisation of asymptotic safety has been advocated
in \cite{Codello:2006in,Niedermaier} where the coefficients of the curvature squared terms approach
an asymptotically free fixed point while a nontrivial fixed point exists for the Newton's
constant and the vacuum energy.
Investigations of this scenario can then be carried out in perturbation theory based on the curvature squared 
couplings.

In order to find meaningful critical theory, which should have a finite number of relevant directions,
we must start from a theory with sufficiently large theory space so that all possible
interactions are taken into account. In practice, to carry out this program, we have to make certain
approximations such that the problem becomes tractable. Finite truncations of couplings are most often
used to study the problem. The idea behind this is that if approximations with finite number of couplings  admit
nontrivial FPs, and these are not significantly affected by further analysis
with more couplings, it would give a support to hypothesis that a continuum limit exists.
Indeed there is accumulating evidence that this is the right direction starting from
Einstein theory with a cosmological constant and additional terms~\cite{lr}--\cite{gies}.
However a problem is that we cannot exhaust all the infinite number of terms within finite truncations of the theory
space. In addition there is also a problem of uniqueness;
how to select the concrete quantum theory from the vast sea of possible terms.

Instead of using finite number of terms, we could consider an action of general function of
the scalar curvature. In this case we may adopt the so-called $f(R)$ approximation
where the effective action takes the form
\bea
\Gamma_k=\int d^d x \sqrt{g} f(R),
\label{action}
\ena
with $k$ indicating the infrared cutoff scale.
Then by considering the flow equation on the maximally symmetric spacetimes, we can reduce the FRGE to
a differential equation for the function $f(R)$ which also depends on the cutoff scale $k$.
The FRGE or flow equation for the functions $f$ were first written in \cite{ms,cpr2} and solved in 
polynomial approximations in a small curvature expansion. These expansions have been extended
up to the 34th power in $R$ \cite{fallslitim} where fixed point solutions were found at each order. 
However due to the existence of some unphysical singularities, these flow equations do not admit global solutions.
These singularities were removed by using a different regularisation scheme in \cite{benedetti} where
the first attempts to find global solutions were made.
It was subsequently shown in \cite{dm1,dm2} that
the flow equations for $f$ written in \cite{ms,cpr2,benedetti}, either
do not have global scaling solutions, or
the solutions are such that all their perturbations are redundant.
More recently flow equations have been written down with a modified functional 
measure \cite{dsz2,dsz3} where global scaling solutions were found.

The first flow equations for $f(R)$ gravity utilised the linear parametrization of the metric 
\bea
g_{\mu\nu}=\bar g_{\mu\nu}+ h_{\mu\nu},
\label{linear}
\ena
between the background $\bar g_{\mu\nu}$ and fluctuation $h_{\mu\nu}$. However this choice is not unique and
it turns out that the flow equations have a simpler form if one uses the exponential
parametrization~\cite{pv}--\cite{opv2}:
\bea
g_{\mu\nu}=\bar g_{\mu\rho}(e^h)^\rho{}_\nu \,.
\label{exp}
\ena
Flow equations using \p{exp} along with a unimodular gauge fixing condition were written down in \cite{opv,opv2}.
Encouragingly, these flow equations admit global scaling solutions, while only differing from the equations
based on the linear parameterisation by terms which vanish on-shell. 
It is particularly striking that for some choices of cutoff, the scaling solutions have an extremely simple,
quadratic form.
These investigations based on the exponential parameterisation are therefore a step forward in the search of
scaling solutions of $f(R)$ gravity.
As noted in \cite{opp}, there are also more advantages of using \p{exp} since it
reduces the dependence of results on the gauge and parameterization (see also \cite{nink}--\cite{Bene2}).
Motivated by these findings we will employ the parametrization~\p{exp} in our study of the flow equations.

The significance of the positive results is reduced by several circumstances.
The first is the restriction of the action to purely background-dependent terms,
the so-called  background field approximation or ``single metric truncation''.
The effective action at finite cutoff cannot be a function of a single metric,
so the classical invariance under the ``shift symmetry''
$\bg_{\mu\nu}\to \bg_{\mu\nu}+\epsilon_{\mu\nu}$, $h_{\mu\nu}\to h_{\mu\nu}-\epsilon_{\mu\nu}$
(in the linear parameterization) is broken.
The dangers of the single-field truncations have been discussed in \cite{lp,manrique}.
One should therefore consider truncations involving either two metrics~\cite{beckerreuter} or
the background metric and a fluctuation field~\cite{cdp,dep,pawlowski} or else solve
the flow equation together with the modified Ward identities of split symmetry~\cite{dm3}.
In any case the scaling solutions found in \cite{opv,opv2} can be at best
an approximation of a genuine scaling solution.
Furthermore, the $f(R)$ approximation is not closed since other curvature invariants will
be generated by the the FRGE on more general spacetimes. In turn these additional interactions
will modify the flow equation for the function $f(R)$.
Hence the approximation assumes these effects are small.

Despite many works in this field, there is little study of the approach on the noncompact or
hyperbolic spaces except for a few works~\cite{Demmel:2014sga,Bene}.
Because the flow equations cannot be directly continued to the spacetimes of
negative curvature, we should study equations also on the hyperbolic spacetimes and
study their properties.
This should also cast some insight to the above problem of the background independence
at the level of topology.
It is thus interesting to study the renormalization group (RG) approach to the gravity
on a hyperbolic space.
In this paper we take a further step in this direction.
We will find that the FRGEs on the hyperbolic spacetimes have quite similar structure to
those on the compact spaces, but there are some differences, which arise from the fact that
some modes in the heat kernel expansion are removed on the compact spaces
but not on the hyperbolic spaces. This seems to cause a problem in the noncompact spaces that
the solutions of polynomial type in the curvature have poor convergence property as the number
of polynomial terms are increased in contrast to the compact case.

Another potential issue was discussed earlier~\cite{opv2,Demmel:2014sga} thought to be related to the compactness
of the background manifold: what is the meaning of coarse-graining on length scales
that are larger than the size of the manifold?
The spectrum on the sphere is characterized by an integer number, which has upper bound depending
on the curvature when we use optimized cutoff~\cite{optimized}, and when the curvature is big enough,
this upper bound becomes too small so that we are not integrating out any modes.
This puts into question the physical meaning of the behaviour of the scaling solution for large $R$
on the compact spaces.
On the other hand, on the hyperbolic space, one might expect that due to its noncompact nature,
the spectrum is continuous without gap and we always integrate out some modes however large or small
the curvature is.
On this ground, it was suggested in \cite{opv2} that the flow equation on noncompact space may not
suffer from this problem.
However, it turns out that there is also a finite gap $\d$ depending on the curvature
in the spectrum even on the hyperbolic space.
Thus we appear again to be faced with the problem of validity for large curvature $|R|$
also in the noncompact spaces. Here we will evaluate the flow equation on a hyperbolic space and reveal
the impact of the finite gap on the flow equation. In fact, we find that all modes can be integrated
out for a finite value of the cutoff due to the gap in the spectrum of the Laplacian, leading already to
the low-energy effective action.

The plan for the rest of this paper is as follows:
In the next section, we set up the flow equation on the hyperbolic spacetimes.
We first consider the equations of motion for $f(R)$ gravity in sect.~2.1. Then in sect.~2.2,
we take the results for the renormalization group equation from \cite{opv}
which takes the same form on hyperbolic or spherical spacetimes and discuss a generalised form.
In sect.~2.3, we discuss the relation of the equation to scalar-tensor theories.
We go on to discuss the structure of the flow equation in sect.~2.4, and derive its concrete
form in four dimensions in sect.~2.5. Approximate forms are given for small curvature in
sect.~2.6 and for critical case in sect.~2.7.
In sect.~3, we discuss solutions in the small curvature approximation, starting with the Einstein-Hilbert
action in sect.~3.1, quadratic ``exact'' solutions in sect.~3.2, polynomial solutions in sect.~3.3.
Global exact solutions are searched for in sect.~4. In particular those characterised by
an asymptotically free $R^2$ coupling are discussed
in sect.~4.1. The asymptotic behavior for global numerical solutions is studied in sect.~4.2 and the possibility
of other global scaling solutions is assessed.
Sect.~5 is devoted to the summary of our results and discussions.
In appendix~\ref{spectrum}, we collect the spectrum and some formulas necessary for the derivation
of the flow equation for the hyperbolic spacetimes in general dimensions,
where we also explicitly show that there is a gap in the spectrum for large curvature $|R|$.
In appendix~\ref{H4vsS4}, we summarize the heat kernel expansions on noncompact spacetimes,
and in particular compare the trace formulas on noncompact and compact spaces for scalars in
subsect.~\ref{H4scalar}, vectors in subsect.~\ref{H4vector} and finally for tensors in
subsect.~\ref{H4tensor}.
Appendix~\ref{sec:flowspect3} contains flow equation for three-dimensional hyperbolic spacetimes
for comparison.

\section{Flow equations for $f(R)$ gravity}
\label{floweq}

In this section, we will discuss the renormalization group flow equations for $f(R)$ gravity
based on an action of the form \p{action} which describes an effective theory where quantum
fluctuations with momentum $p^2 > k^2$ included. For finite $k$, the effective action 
has an infrared (IR) cutoff which suppresses both the UV and IR divergences in the FRGE.
In the limit of $k \to 0$, the cutoff is removed and we obtain the full quantum effective
action $\Gamma = \Gamma_0$.

\subsection{Equations of motion}
Varying \p{action} with respect to the metric, we obtain the $k$-dependent equation of motion
\be \label{EofMfull}
\frac{1}{2} f(R) g^{\mu \nu} - f'(R) R^{\mu \nu}  + (g^{\lambda \nu} g^{\rho \mu}
-g^{\mu \nu} g^{\rho \lambda}) \nabla_\lambda \nabla_\rho f'(R) = 0,
\ee
which takes the same form as the classical $f(R)$ equation of motion.
Here we will be interested in the background metrics $\bar{g}_{\mu\nu}$ which describe maximally
symmetric spacetimes with the Riemann curvature given by 
\be
\bar{R}_{\mu \nu \rho \sigma} = \frac{\bar{R}}{d(d-1)} (\bar{g}_{\mu \rho} \bar{g}_{\nu \sigma}
- \bar{g}_{\mu \sigma} \bar{g}_{\nu \rho}),
\ee
with $\bar{R}$ a constant over spacetime.
We note that this reduces any tensor structure depending on the Riemann tensor
to a function of the curvature scalar.
For such backgrounds, the equation of motion reduces to the form 
\be \label{EofM}
\frac{d}{2} f(\bar{R}_0) - R_0 f'(\bar{R}_0) = 0 \,,
\ee 
which is simply a constraint on the scalar curvature $\bar{R}=\bar{R}_0$.
For values of $\bar{R} = \bar{R}_0$ which satisfy \p{EofM}, we can say that the equations of motion
admit constant curvature solutions corresponding to de Sitter or anti-de Sitter depending on
the sign of $\bar{R}_0$. For the special case in which $\bar{R}_0 = 0$ satisfies \p{EofM},
flat space is a solution to the equation of motion.

\subsection{Functional renormalization group equations for $f(R)$ gravity in the exponential
parameterisation}
\label{sec:frg}

Here we follow \cite{opv,opv2} to derive a flow equation in the $f(R)$ approximation to which
we refer for further details.
We use the exponential parameterisation of the metric
\bea
g_{\mu\nu} &=& \bg_{\mu\rho} ( e^h )^\rho{}_\nu\
= \bg_{\mu\nu} + h_{\mu\nu} + \frac12 h_{\mu\la} h^\la{}_\nu + \ldots,
\label{nonlinear}
\ena
and partial gauge fixing $\bar{g}^{\mu\nu} h_{\mu\nu} = 0$ corresponding to a unimodular gauge.
Additional gauge fixing is then also required to fix the remaining $d-1$ diffeomorphisms.
After this is done, the action \p{action} will be supplemented with additional terms depending on
ghost fields.
The main input for the flow equation then comes from the second functional derivative of
the action with respect to the (gauge fixed) metric fluctuations and the ghost fields. 
Then, following the standard methods, we arrive at the  FRGE~\cite{opv,opv2}
\bea
\dot \G_k \hs{-2}&=&\hs{-2}\frac{1}{2} \mbox{Tr}_{(2)}
\left[\frac{\dot f'(\br) R_k(\Delta-\alpha\bR)
+f'(\br) \dot R_k(\Delta-\alpha\bR)}{f'(\br) \left(P_k(\Delta-\alpha\bR)
+\alpha\bR+\frac{2}{d(d-1)}\br \right)}\right]
\nn &&
- \frac{1}{2} \mbox{Tr}_{(1)}\left[ \frac{\dot R_k(\Delta-\gamma\bR)}{P_k(\Delta-\gamma\bR)
+\gamma\bR
-\frac{1}{d}\br} \right]
\nn &&
+\; \frac{1}{2} \mbox{Tr}_{(0)} \left[ \frac{\dot f''(\br)R_k(\Delta-\beta\bR)
+f''(\br) \dot R_k(\Delta-\beta\bR)}
{f''(\br) \left(P_k(\Delta-\beta\bR)
+\beta\bR-\frac{1}{d-1}\br \right)+\frac{d-2}{2(d-1)}f'(\br)} \right]
 ,~~~
\label{frge}
\ena
where the dot denotes derivative with respect to the RG time $t= \log k/k_0$ (with $k_0$ an arbitrary reference scale),
$P_k(z) = z +R_k(z)$
with the cutoff function $R_k(z)$ and $\Delta =-\nabla^2$ is the Laplacian.
The subscripts on the traces represent contributions from different spin sectors;
$(2)$ denoting a trace over transverse-traceless symmetric tensor modes,
$(1)$ a trace over transverse-vector modes and
$(0)$ a trace over scalar modes.
Here $\a$, $\b$ and $\c$ are free parameters the choice of which corresponds to
the choice of RG schemes along with the choice of the function $R_k(z)$.
We note that the traces can in principle be evaluated for both negative and positive curvatures
and in any dimension $d$. Here we study the case of $\bar{R} < 0$ in $d=4$ dimensions for which
we need the spectrum of the Laplacian on a symmetric hyperbolic space $H^4$. We give
the necessary formulas in Appendix~\ref{spectrum}.

To understand the structure of this equation, we first note that it describes graviton
fluctuations, comprised of the transverse-traceless fluctuations and the ghost transverse vector
fluctuations, plus a scalar fluctuation, the `scalaron', which is absent in the case $f''(R) =0$. 
We can then understand the terms which arise in the RHS of the flow
equation proportional to $\dot{f}'(\bR)$ and $\dot{f}''(\bR)$ as curvature-dependent anomalous
dimensions
\be \label{etas}
\eta_{2}(\bR) = - \frac{\dot{f}'(\bR)}{f'(\bR)}\,,\,\,\,\,\,  \eta_{0}(\bR)
 = - \frac{\dot{f}''(\bR)}{f''(\bR)}\,,\,\,\,\,\,\,\, \eta_{1}(\bR) = 0 \,.
\ee
These particular choices for the anomalous dimension come from the background field approximation.
We note that approximations that go beyond the background field could instead determine $\eta_{j}$
from the flow of the propagator which can in general lead to momentum and curvature dependencies as
$\eta_{j}= \eta_{j}(\bR,\Delta)$.
Then a more general equation can be written as 
\bea \label{generalised_flow}
\dot \G_k \hs{-2}&=& \hs{-2}\frac{1}{2} \mbox{Tr}_{(2)}
\left[\frac{\dot R_k(\Delta-\alpha\bR)-\eta_{2}(\bR,\Delta) R_k(\Delta-\alpha\bR)}{P_k(\Delta-\alpha\bR)
+\alpha\bR+\frac{2}{d(d-1)}\br }\right] \nn
&-& \hs{-2}\frac{1}{2} \mbox{Tr}_{(1)}
\left[\frac{\dot R_k(\Delta-\gamma\bR)-\eta_{1}(\bR,\Delta) R_k(\Delta-\gamma\bR)}{P_k(\Delta-\gamma\bR)
+\gamma\bR-\frac{1}{d}\br}\right] \nn
&+& \hs{-2}\frac{1}{2} \mbox{Tr}_{(0)}
\left[\frac{\dot R_k(\Delta-\beta\bR)-\eta_{0}(\bR,\Delta) R_k(\Delta-\beta \bR)}{P_k(\Delta-\beta \bR)
+\beta \bR + \frac{1}{f''(\br)} \left(\frac{d-2}{2(d-1)}f'(\br) -\frac{f''(\br)}{d-1}\br\right)}\right].
\ena
This flow equation is equal to Eq.~\p{frge} only when the anomalous dimensions are given
by Eq.~\p{etas}.
Taking instead $\eta_j = 0$, we recover the one-loop type RG equation.
In the background field approximation, the ghosts have a vanishing anomalous dimension~\p{etas}
which means their contribution is effectively one loop. Motivated by the observation that only
a combination of the transverse-traceless modes and the transverse-vector ghosts produce
the $d(d-3)/2$ physical fluctuations of the graviton, an alternative choice is to identify
the anomalous dimension of the ghosts with the graviton:
\be \label{etas2}
 \eta_{1}(\bR) = \eta_{2}(\bR) = - \frac{\dot{f}'(\bR)}{f'(\bR)}\,,\,\,\,\,\, 
\eta_{0}(\bR) = - \frac{\dot{f}''(\bR)}{f''(\bR)}\,.
\ee
At the level of the Einstein-Hilbert truncation, this approximation \cite{falls} leads to
a critical exponent which is in good agreement with lattice studies \cite{Hamber}.
Here we study the standard
background field approximation \p{etas} as well as the one-loop approximation $\eta_j = 0$.

\subsection{Relation to scalar-tensor theories}

Since classically $f(R)$ gravity is equivalent to Einstein gravity coupled to a scalar theory
(see e.g. \cite{Sotiriou:2008rp}),
it is interesting to see how much of this equivalence is carried over in the structure of
the RG equation. Motivated by this relation, the RG flow of Brans-Dicke theory has been studied
in \cite{Benedetti:2013nya}.

First we note that Einstein gravity is obtained when we take $f(R)$ to be a linear
function leading to the absence of spin zero trace since $f''(\bR) = 0$.
This is to be expected since in Einstein gravity there is no scalar degree of freedom.
Let us then consider a scalar field with an action
\be
S_{\phi} = \int d^dx \sqrt{g} \left[ \frac{1}{2} \nabla_\mu \phi \nabla^\mu \phi + U(\phi) \right] .
\ee
This would lead to a trace in the flow equation of the form
\be \label{scalar_trace}
\hs{-2}\frac{1}{2} \mbox{Tr}_{(0)}
\left[\frac{\dot R_k(\Delta-\beta\bR)- \eta_{0}(\bR,\Delta) R_k(\Delta-\beta \bR)}{P_k(\Delta-\beta \bR)
+\beta \bR  +  U''(\phi)      }\right],
\ee
where we have allowed for a general curvature- and momentum-dependent anomalous dimension
and have taken $\phi$ to be a spacetime constant.
Comparing \p{scalar_trace} with \eq{generalised_flow}, we see that structurally this term is
equal to the scalar trace in the $f(R)$ theory with a replacement
\be \label{U2}
U''(\phi)  \to \frac{1}{f''(\br)} \left(\frac{d-2}{2(d-1)}f'(\br)  - \frac{f''(\br)}{d-1}\br\right) .
\ee
One can now check the consistency of the above relation by noting that
via field redefinitions, which equate $f(R)$ gravity to a scalar-tensor theory
in the Einstein frame, the field $\phi$ is identified as
\be
\phi = - f'(\bR) \,.
\label{ide}
\ee
To obtain an expression for $U''(\phi)$, we follow \cite{Hindmarsh:2012rc} taking the trace of
the $f(R)$ equation of motion \p{EofMfull} and compare it to the scalar fields equation of motion
\be
- \Delta f'(\bR)  + \frac{1}{d-1} \left( \bR f'(\bR) -  \frac{d}{2} f(\bR) \right) = 0~~~
\Longleftrightarrow~~~
  \Delta \phi  + U'(\phi) = 0,
\ee
which allows us to identify
\be
U'(\phi) = \frac{1}{d-1} \left(  \br f'(\br) - \frac{d}{2} f(\br) \right) .
\ee
Taking one further derivative with respect to $\phi$, using
$\frac{d}{d\phi}=\frac{d\br}{d\phi}\frac{d}{d\br}=-\frac{1}{f''(\br)}\frac{d}{d\br}$ from \p{ide},
we obtain the second derivative of the potential:
\be \label{U''}
U''(\phi) =   \frac{1}{f''(\br)}   \left( \frac{d-2}{2(d-1)} f'(\br) - \frac{\br}{d-1} f''(\br)\right),
\ee
in agreement with Eq.~\p{U2}.
Thus, we see that the term indeed represents the contribution of the scalar component of the metric.
Despite this, it is clear that on the background level the flow equations for scalar-tensor theories
and $f(R)$ gravity will not be related by a simple change of variables since the $f(R)$ flow equations
depend only on the scalar curvature $R$ whereas in scalar-tensor theory
the scalar field would introduce a further variable $\phi$.
Nonetheless the observations made here may be of use to further understanding the relation of
RG flows for scalar-tensor
theories~\cite{pv,LPV,Benedetti:2013nya,Narain:2009fy,Narain:2009gb,Henz:2013oxa,Henz:2016aoh,Saltas:2015vsc}
to that of pure gravity.

\subsection{Cutoffs and dimensionless flow equation}
\label{sec:cutoff}

In this paper, we will utilise the optimised cutoff function $R_k = (k^2 - z) \Theta(k^2 -z)$
where $\Theta(k^2 -z)$ is the Heaviside theta function.
In order to guarantee that the operators $\Delta-\alpha\bR$, $\Delta-\beta\bR$,
$\Delta-\gamma\bR$ have positive spectrum, the parameters $\alpha$, $\beta$ and $\gamma$
should satisfy certain bounds.
The spectrum of $\Delta$ on $H^d$ is given in Appendix~\ref{spectrum}.
Positivity then requires that
\bea
\alpha > -\frac{(d-1)^2+8}{4d(d-1)}\ ;\qquad
\gamma > -\frac{(d-1)^2+4}{4d(d-1)}\ ;\qquad
\beta > -\frac{(d-1)^2}{4d(d-1)}\ ,
\label{zmb}
\ena
where if these inequalities are not satisfied, some modes will not be integrated out even
in the limit $k \to 0$.
Note that the scalar curvature is negative $\br<0$ in deriving this result.

As usual, we now introduce the dimensionless quantities
\bea
r =\br k^{-2},~~~
\vp(r) = k^{-d} f(\br),
\label{diml}
\ena
so that $\dot f(\br)=k^d[\dot\vp(r)-2r\vp'(r) + d\vp(r)]$, 
$\dot f'(\br)=k^{d-2}[\dot\vp'(r)-2r\vp''(r) + (d-2)\vp'(r)],$
and $\dot f''(\br)=k^{d-4}[\dot\vp''(r)-2r\vp'''(r) + (d-4)\vp''(r)]$.
In dimensionless form, the flow equation can be written as
\bea \label{fRflow}
\dot{\varphi}(r) + d \varphi(r) - 2 r \varphi'(r)   &=&  \frac{N_2(r)}{D_2(r)}
-  \frac{N_1(r)}{D_1(r)}  +  \frac{N_0(r)}{D_0(r,\varphi', \varphi'')}
+ \eta_2(r,\varphi', \varphi'', \dot{\varphi}')   \frac{\tilde{N}_2(r)}{D_2(r)} 
\nonumber \\
&+& \eta_0(r,\varphi', \varphi'', \varphi''',\dot{\varphi}'')
 \frac{\tilde{N}_0(r)}{D_0(r,\varphi', \varphi'')}
\ .
\ena
Here we have the denominators
\bea
D_0 &=& 1 + \frac{d-2}{2(d-1)} \frac{\varphi'(r)}{\varphi''(r)} - \frac{r}{d-1} + \beta \, r, \\
D_1 &=& 1  - \frac{1}{d} r  + \gamma \, r,  \\
D_2 &=&  1  +  \frac{2}{d(d-1)} r  + \alpha \, r ,
\ena
and the numerators which are traces
\bea \label{numerators}
N_j(r) =  v^{-1} \Tr_{(j)} \left[ \Theta(1-y + \alpha_j \, r)    \right], \,\,
\tilde{N}_j(r) =  \frac{1}{2} v^{-1} \Tr_{(j)} \left[(1-y + \alpha_j \, r)
\Theta(1 -y  + \alpha_j \, r)\right]  ,\,\,
\ena
where $\alpha_j =\{ \beta, \gamma, \alpha\}$ ($j=0,1,2$ in that order) are the endomorphism parameters,
$y= \Delta k^{-2}$ and $v = k^d \int d^dx \sqrt{\bar{g}}$ is the dimensionless volume.
Here we note that the denominators $D \propto \Gamma^{(2)}_k+ R_k > 0$
must be positive definite to ensure the convexity of the effective action.
The anomalous dimensions of the scalaron and graviton defined in \p{etas} are expressed
in dimensionless form as
\be
\eta_2 = \frac{ \dot{\varphi}'(r) +(d-2) \varphi'(r) - 2 r \varphi''(r)}{\varphi'(r)}
  \,,  \,\,\,\,\,\,\,\, \,\,\,\,\,\,\,\,
\eta_0 =   \frac{\dot{\varphi}''(r) + (d-4) \varphi''(r) - 2 r \varphi'''(r)}{\varphi''(r)} \,.
\ee
A scaling solution, or fixed point solution, to \p{fRflow} is a solution where $\dot{\vp}$
 and its derivatives with respect to $r$ vanish. 
Looking at the LHS of the flow equation $\eq{fRflow}$, one observes that for values
$r_0= R_0/k^2$ satisfying \p{EofM} the second and third terms vanish leaving only $\dot{\varphi}(r)$.
It follows that for a fixed point solution, the vanishing of the RHS
of the equation, at some value of the curvature $r = r_0$, implies a constant curvature solution
to the equation of motion \p{EofMfull}. It has been argued \cite{dm2} that without such solutions,
all eigen-perturbations around the fixed point are redundant.

\subsection{Spectral sum approach for $d=4$}
\label{sec:flowspect}

We can compute the traces in Eq.~(\ref{frge}) by integrating the corresponding functions
where the eigenvalues and spectral measure are given in Appendix~\ref{spectrum}.
On the hyperbolic space $H^4$,
we have
\be
N_j(r)  =
\frac{2(2j+1)}{\pi^3} \Big(\frac{-\br}{12}\Big)^2
\int_0^{\bar\la^{(j)}} d\la \frac{\pi[\la^2+\left(j+ 1/2\right)^2]}{16} \la \tanh(\pi\la),
\ee
and
\be
\tilde{N}_j(r)  =
\frac{1}{2} \cdot \frac{2(2j+1)}{\pi^3} \Big(\frac{-\br}{12}\Big)^2
\int_0^{\bar\la^{(j)}} d\la \frac{\pi[\la^2+\left(j+ 1/2\right)^2]}{16} \la \tanh(\pi\la)
\left(1 +\frac{\la^2+\frac{9}{4}+j}{12} r + \alpha_i \, r\right),
\ee
Due to the theta functions, the integrals over $\lambda$ are cutoff at
\bea
\bar{\la}^{(2)}\!=\sqrt{\frac{12}{-r}\!-\frac{17}{4}-12\a} \,, \quad
\bar{\la}^{(1)}\!=\sqrt{\frac{12}{-r}\!-\frac{13}{4}-12\c} \,, \quad
\bar{\la}^{(0)}\!=\sqrt{\frac{12}{-r}\!-\frac{9}{4}-12\b} \,. \quad
\label{upperlimits}
\ena
An analogous upper bound is found
on the sphere~\cite{opv2} where the spectrum sum has a maximum ``angular momentum'' $\ell$.
In both cases the upper bound exists because there is a finite gap $\delta_j$ in the
eigen-spectrum of the operators corresponding to the smallest eigenvalue, see appendix~\eq{gap}. 
Explicitly for $d=4$ and on the hyperbolic space we have the smallest eigenvalue
\be \label{gap4}
\delta_j = -\frac{\frac{9}{4}+j}{12} \br  - \alpha_j \br ,
\ee
for the operators $-\nabla^2 - \alpha_j \br$ acting on a field of spin $j$.
As a consequence of the upper bounds \p{upperlimits}, the traces $N_j$ and $\tilde{N}_j$ have support
only for curvatures in the range
\bea
r_{{\rm crit},j} \equiv \left\{- \frac{1}{\b+3/16},- \frac{1}{\c+13/48},- \frac{1}{\a+17/48}
\right\} \leq r \leq 0 ,
\label{range}
\ena
for which they are positive $N_j > 0$ and $\tilde{N}_j > 0$ and are identically  zero outside this range. 
The meaning of this is that once $r<r_{{\rm crit},j}$, all modes of spin $j$ have been integrated out
in the functional integral. A similar critical value $r_{{\rm crit},j}$ of $r$ exists on the four-sphere \cite{dsz3} for which all modes
 are integrated when $r>r_{{\rm crit},j}$. 
On the hyperbolic space, it is only by fixing the endomorphism parameters to the critical values 
\bea
\a = -\frac{17}{48},~~
\b = -\frac{9}{48},~~
\c = -\frac{13}{48},
\label{cut3}
\ena
that the range \p{range} extends to  $r_{{\rm crit},j}= - \infty$, whereby the gap
in the eigen-spectrum \p{gap4} vanishes for each spin $j=0,1,2$. Conversely if none of
the parameters are given by \p{cut3}, then the RHS of the flow equation will vanish
once $r < r_{{\rm crit}, j}$ for all $j$.

Within the range \p{range}, we have
\bea
N_0(r) &=& \frac{r^2}{4608 \pi^2} ( I_1(\bar\la^{(0)}) + 4 I_3(\bar\la^{(0)}))  \,, \nn
N_1(r) &=& \frac{r^2}{1536 \pi^2} ( 9 I_1(\bar\la^{(1)}) + 4 I_3(\bar\la^{(1)}))  \,, \nn
N_2(r) &=& \frac{5 r^2}{4608 \pi^2} ( 25 I_1(\bar\la^{(2)}) + 4 I_3(\bar\la^{(2)}))  \,, \nn
\tilde{N}_0(r) &=& \frac{r^2}{384 \pi^2} \frac{ ((48 \beta +9) r+48)  I_1(\bar\la^{(0)})
+ (192 \beta  r+40 r+192) I_3(\bar\la^{(0)})+2 r\,I_5(\bar\la^{(0)}) }{1152}\,, \nn
 \tilde{N}_2(r) &=&   \frac{r^2}{384 \pi^2}  \frac{125 ((48 \alpha +17) r+48) I_1
\left(\bar{\lambda }^2\right)+40 \left(3 (8 \alpha  r+7 r+8)
I_3\left(\bar{\lambda }^2\right)+2 r I_5\left(\bar{\lambda }^2\right)\right)}{1152} \,. \nn
\label{tracesr}
\ena
where the functions $I_n(x)$ are defined by
\bea 
I_n(x) = \int_0^x \tilde{x}^n \tanh(\pi \tilde{x}) d\tilde{x}.
\label{integral}
\ena
The traces~\p{tracesr} can be evaluated as functions
of $r$ for fixed values of the endomorphisms.
Evaluating the traces for $r > r_{{\rm crit},j}$, we obtain the following flow equation:
\bea
&& \hs{-10}
\dot \vp -2r \vp'+4\vp \nn
&=& \frac{c_1 (\dot\vp'-2r \vp'')+c_2 \vp'}{\vp'[6+(6\a+1)r ]}
+ \frac{c_3 (\dot\vp''-2r \vp''')+c_4 \vp''}{[3+(3\b-1)r]\vp''+\vp'}
- \frac{c_5}{4+(4\c-1) r}.
\label{erge}
\ena
where the coefficients $c_i$ depend on the scalar curvature $r$.
Notice the similarity of the structure of these equations to those on sphere~\cite{opv,opv2}.
Here we note the relations $c_1 = 6  \tilde{N}_2(r)$, $c_2 = 6  N_2(r) + 126  \tilde{N}_2$,
$c_3 = 3 \tilde{N}_0$, $c_4 = 3 N_0$ and $c_5= 4 N_1$ between the coefficients and the traces.
As a result we note that each coefficient is $c_i \geq 0$ due to the (semi)-positivity of the traces.
The coefficients $c$ are given by
\bea
c_1 &=& \frac{5r^2}{384\pi^2} \Big[ \frac{r}{12} I_5(\bar\la^{(2)})
+ \Big(1+\Big(\a+\frac{7}{8}\Big)r\Big) I_3(\bar\la^{(2)})
+ \frac{25}{4} \Big(1+\Big(\a+\frac{17}{48}\Big)r\Big) I_1(\bar\la^{(2)})\Big], \nn
c_2 &=& \frac{5r^2}{384\pi^2} \Big[ \frac{r}{6} I_5(\bar\la^{(2)})
+ 2 \Big(2+\Big(\a+\frac{7}{8}\Big)r\Big) I_3(\bar\la^{(2)})
+ \frac{25}{2} \Big(2+\Big(\a+\frac{17}{48}\Big)r\Big) I_1(\bar\la^{(2)})\Big], \nn
c_3 &=& \frac{r^2}{768\pi^2} \Big[ \frac{r}{12} I_5(\bar\la^{(0)})
+ \Big(1+\Big(\b+\frac{5}{24}\Big)r\Big) I_3(\bar\la^{(0)})
+ \frac{1}{4} \Big(1+\Big(\b+\frac{3}{16}\Big)r\Big) I_1(\bar\la^{(0)})\Big], \nn
c_4 &=& \frac{r^2}{384\pi^2} \Big[ I_3(\bar\la^{(0)}) + \frac{1}{4} I_1(\bar\la^{(0)})\Big], \nn
c_5 &=& \frac{r^2}{96\pi^2} \Big[ I_3(\bar\la^{(1)}) + \frac{9}{4} I_1(\bar\la^{(1)})\Big].
\label{coe_spectral}
\ena

We can explicitly evaluate the integrals $I_n(x)$ for $x>0$:
\bea
I_1(x>0) &=& \frac{x^2}{2} + \frac{x}{\pi} \ln(1+e^{-2\pi x})
-\frac{1}{2\pi^2} \mbox{\,Li}_2(-e^{-2\pi x}) -\frac{1}{24}, \nn
I_3(x>0) &=& 
 \frac{x^4}{4} +\frac{x^3}{\pi} \ln(1+e^{-2\pi x})
-\frac{3}{4\pi^4} \Big[ 2\pi^2 x^2 \mbox{\,Li}_2(-e^{-2\pi x})
- 2\pi x \mbox{\,Li}_3(-e^{-2\pi x}) \nn
&& - \mbox{\,Li}_4(-e^{-2\pi x}) \Big]
-\frac{7}{960}, \nn
I_5(x>0) &=& 
 \frac{x^6}{6} +\frac{x^5}{\pi} \ln(1+e^{-2\pi x})
-\frac{5}{4\pi^6} \Big[ 2\pi^4 x^4 \mbox{\,Li}_2(-e^{-2\pi x})
+4 \pi^3 x^3 \mbox{\,Li}_3(-e^{-2\pi x}) \nn
&& +6\pi^2 x^2 \mbox{\,Li}_4(-e^{-2\pi x})
+6 \pi x \mbox{\,Li}_5(-e^{-2\pi x}) +3 \mbox{\,Li}_6(-e^{-2\pi x}) \Big] -\frac{31}{8064},
\ena
where
\bea
\mbox{Li}_n(x) = \sum_{k=1}^\infty \frac{x^k}{k^n},
\ena
is the polylogarithm. When $x\leq 0$ the integrals vanish identically
\be
I_1(x \leq	0) = 0  \,, \,\,\,\,\, I_3(x\leq0) = 0  \,, \,\,\,\,\, I_5(x\leq0) = 0 
\ee  
In this way we have obtained an explicit form for the flow equation on $H^4$.
We stress that this explicit form is unlike that obtained on $S^4$ with the optimised cutoff where
additional approximations are made to smoothen the functions obtained by evaluating the traces.
In particular the property that the traces should vanish when the last mode is integrated out is lost 
and thus the proper IR limit cannot be reached. In contrast, here since we evaluate the traces directly,
the IR limit is obtained once $\bar{R}/k^2 < r_{{\rm crit}, j}$ (for all spins $j$) whereby $R_k = 0$ and we 
obtain already the full effective action $\Gamma_k = \Gamma_0$.

\subsection{Small curvature expansion}

For small curvature $|r|$, we expect that the early-time heat kernel expansion should also
provide an accurate evaluation of the traces appearing in the flow equation.
Here we note that for small $-r$, each of the $\bar{\la}^{(j)}$ diverges and hence
the exponentials occurring in $I_n(\bar{\la}^{(j)})$ will be small.
It then follows that we can make the following approximations:
\bea
I_1(x) &\approx& \frac{x^2}{2}  -\frac{1}{24}, \nn
I_3(x) &\approx& 
 \frac{x^4}{4}
-\frac{7}{960}, \nn
I_5(x) &\approx& 
 \frac{x^6}{6}  -\frac{31}{8064}.
\ena
Under this approximation, the coefficient $c_i$ are then simply polynomials in the curvature:
\bea
c_1 &\approx& \frac{1}{32\pi^2} \Big[ \left(5 \alpha ^3-\frac{5 \alpha ^2}{2}-\frac{271 \alpha}{72}
   -\frac{7249}{9072}\right) r^3+\left(15 \alpha ^2-5 \alpha
   -\frac{271}{72}\right) r^2+\frac{5}{2} (6 \alpha -1) r+5\Big], \nn
c_2 &\approx& \frac{1}{32\pi^2} \Big[ \left(10 \alpha ^3-5 \alpha ^2-\frac{271 \alpha }{36}
   -\frac{7249}{4536}\right)r^3+\left(60 \alpha ^2-20 \alpha -\frac{271}{18}\right) r^2
   +15 (6 \alpha -1) r+40 \Big], \nn
c_3 &\approx& \frac{1}{32\pi^2} \Big[\left(\frac{1}{2} \b^3+\frac14 \b^2+\frac{29 \b}{720}
   + \frac{37}{18144}\right) r^3 + \frac12 \left( 3 \b^2+  \b + \frac{29}{360} \right)r^2
   + \frac{1}{4} (6 \b + 1) r + \frac{1}{2} \Big], \nn
c_4 &\approx& \frac{1}{32\pi^2} \Big[ \left(3 \b^2+ \b +\frac{29}{360}\right) r^2+ (6 \b+1)r
   +3 \Big], \nn
c_5 &\approx& \frac{1}{32\pi^2} \Big[ \left(12 \gamma ^2+2 \gamma -\frac{67}{180}\right) r^2
   + 2(12 \gamma +1) r+12\Big].
\label{coe_smallr}
\ena
Here we have derived this approximation which holds in the asymptotic limit $-r \to 0$
of the full spectral form of the traces. One can confirm that by using the early-time heat
kernel expansion, we obtain the same coefficients~(\ref{coe_smallr}).
These coefficients differ only slightly from those obtained using the heat kernel expansion on the sphere.
To account for the difference between topologies, we should separate out the contributions
(which must be added or subtracted appropriately) from
the $\Sigma_{KV}=10$ Killing vectors (KVs), the $\Sigma_{CKV}=5$ conformal Killing vectors (CKVs)
and the single constant mode $\Sigma_{\rm const.}=1$. These are present on the sphere but not
on the hyperbolic space since it is non-compact. Details can be found in Appendix~\ref{H4vsS4}.

\begin{figure}
\centering
\includegraphics[width=0.8\textwidth]{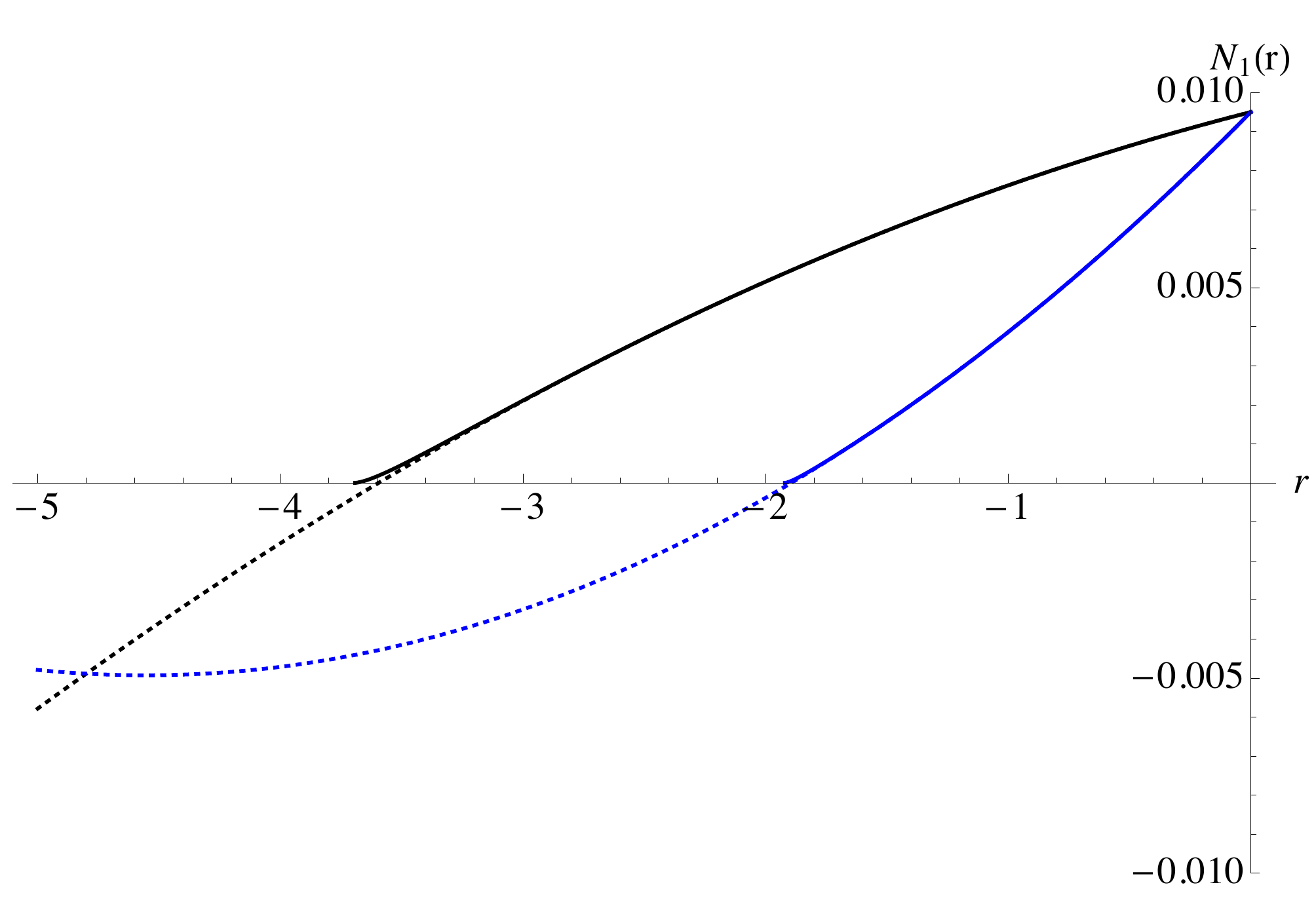}
\caption{Lower (blue) line is for $\gamma = 1/4$ and upper (black) line is for $\gamma = 0$.
Dotted lines are the small $-r$ expansion and the solid lines is the exact trace. }
\label{N1}
\end{figure}

\begin{figure}
\centering
\includegraphics[width=0.8\textwidth]{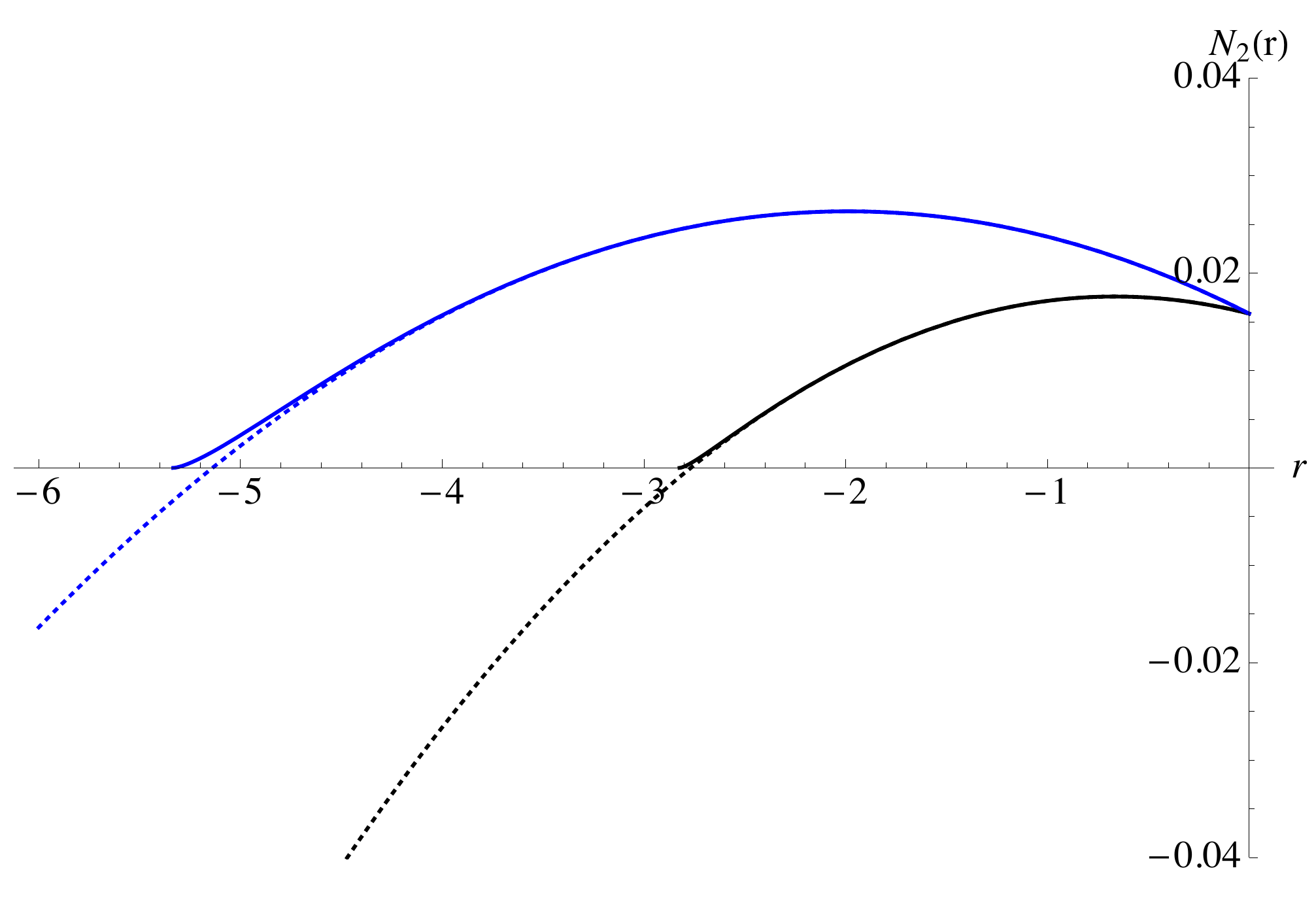}
\caption{Upper (blue) line is for $\a = -1/6$ and lower (black) line is for $\a = 0$.
Dotted lines are the small $-r$ expansion and the solid lines is the exact trace.}
\label{N2}
\end{figure}

Interestingly, the small $|r|$ approximation is very close to the exact result even for
$-r > 1$
and only breaks down as $r$ approaches $r_{{\rm crit},j}$.
To see this, we plot the $N_1$ traces in Fig.~\ref{N1} and $N_2$ traces in
Fig.~\ref{N2} for different values of the endomorphisms.
In consequence we can use these approximate expressions to find solutions to our flow equation
even away from the $-r \to 0$ limit.

\subsection{Expansion around the critical values}

It is interesting to see the $x \to 0$ limit of the integrals $I_n(x)$ defined by~\p{integral}
corresponding to the IR limit.
In this limit, we find
\bea
I_1(x) &\approx& \frac{\pi x^3}{3} \Big( 1 - \frac{(\pi x)^2}{5} + \frac{2}{35} (\pi x)^4
+ \cdots \Big),
\nn
I_3(x) &\approx& \frac{\pi x^5}{5} \Big( 1 - \frac{5}{21}(\pi x)^2 + \frac{2}{27} (\pi x)^4
+ \cdots \Big),
\nn
I_5(x) &\approx& \frac{\pi x^7}{7} \Big( 1 - \frac{7}{27}(\pi x)^2 + \frac{14}{165} (\pi x)^4
+ \cdots \Big).
\label{criticalexp}
\ena
This approximation is valid when the curvature is close to the boundary of the range~\p{range},
and it appears that we get non-analytic behaviour due to the square roots in \p{upperlimits}.
However this non-analyticity does not lead to a complex  flow equation since \p{criticalexp} is used only
in the range of $r$~\p{range} where the square root is real, and the term vanishes beyond
the range.
A consequence of the non-analytic behaviour will be that the solution to the fixed point equation
will generically be non-analytic around the points where $r =r_{{\rm crit}, j}$ for which $x=0$.  
Similar behaviour is observed for the odd dimensional case, where the integrals does not involve
arctangent so that they give exact polynomials. This can be seen in the the exact expression
for $d=3$ given in appendix~\ref{sec:flowspect3}.

We note that the expansions \p{criticalexp} correspond to asymptotically large negative $r$ 
for the choice of endomorphism \p{cut3} 
whereby the range \p{range} extends to  $r= - \infty$,
and the large $r$ corresponds to small $x$.

\section{Polynomial solutions in four dimensions}
\label{sec:solspect}

We now analyse the FP solutions $\dot \vp = 0$ of the flow equation for $d=4$ in the small $|r|$
approximation \p{coe_smallr}.
As we have noted, the flow equation then takes a similar form to that of the equation on
a sphere using the heat kernel expansion.
The differences arise from the difference in the treatment of some modes in the heat kernel
expansions for different topologies.
In addition, since we know the exact equation we can better assess the approximation being made. 

\subsection{Einstein-Hilbert at one-loop}

It is well known that pure gravity with the Einstein-Hilbert action is renormalizable on shell
at one-loop. Furthermore it is found that there is a UV FP within this approximation
when we ignore higher order curvature terms and truncate the theory to just the Einstein-Hilbert action.
It is interesting therefore to see what is modified if we do not neglect the higher order
curvature terms which are present in the $f(R)$ approximation.
An important point is that one-loop renormalization in Einstein theory relies on the fact
that the Gauss-Bonnet term $\int d^4x E\;[\, \equiv \int d^4x\, (R_{\mu\nu\rho\la}^2-4 R_{\mu\nu}^2+R^2)]$
is a topological invariant.
On a constant curvature spacetime, we have that $\int d^4x E \propto  \int d^4x R^2$
and hence we can neglect terms which are of order $R^2$ in the flow equation by including,
in addition to the Einstein-Hilbert action, also the Gauss-Bonnet term.
We therefore look for a FP solution to the one-loop $f(R)$ flow equation (i.e. where $\eta =0$) with
\be
\vp(r) = g_0 + g_1 r,
\ee
inserted on the RHS of the one-loop equation and with an additional term $ - \frac{r^2}{\sigma}$
which appears only on the LHS originating from the topological invariant.
The couplings $g_0 = \lambda/(8\pi g)$ and $g_1 = -1/(16 \pi g)$ are then related to
the dimensionless Newton's coupling $g = k^2 G$ and cosmological constant $\lambda = k^{-2} \Lambda$
which form the dimensionless product $G \Lambda = g \lambda$.
It is evident that such a solution, containing only two dynamical couplings, cannot be global.
However if we use the asymptotic approximation~\p{coe_smallr}, we can still hope that no terms
outside Einstein-Hilbert truncation are renormalized in the small curvature limit.
In this limit one finds
\be
4 g_0 + 2 r g_1 + \beta_\sigma r^2/\sigma^2 =  \frac{\frac{1}{2} r \left(900 \alpha -540 \gamma
-\frac{905 r}{6 \alpha r+r+6}-\frac{158 r}{4 \gamma  r-r+4}-675\right)+180}{2880 \pi^2},
\ee
where we are looking for a FP for $g_0$ and $g_1$. One observes that
for general $\alpha$ and $\gamma$, there is no solution if we require that this be valid
for arbitrary $r$. Instead one is forced to set $\alpha = -1/6$ and $\gamma = 1/4$ such that
the RHS is quadratic in $r$. Then we obtain the FP
\be
g_0 = \frac{1}{64 \pi^2} \,,   \,\,\,\,\,\,   g_1 = - \frac{1}{12 \pi^2}  \,.
\ee
Additionally one finds that
\be
\beta_\sigma = - \frac{571}{45} \frac{1}{384 \pi^2} \sigma^2 ,
\ee
which is actually a universal result independent of the regulator function and endomorphisms
as can be confirmed by comparison with the well known one-loop analysis \cite{Christensen}.
Here observe that the requirement that there be no nonzero irrelevant couplings
in the small curvature limit actually specifies the values of the endomorphisms.
The same can be said of the shape function since it is only by using the optimised cutoff
that the regulated Hessians can become constants independent of $\br$.
This singles out this choice of regulator up to rescaling of the cutoff scale which keeps
the dimensionless product
\be
g_* \lambda_* = \frac{9 \pi}{128},
\ee
invariant.
Given that we have an exact scaling solution in the limit $-r \to 0$,
we could in principle use this information to set the boundary condition and integrate the FP
equation towards $r \to - \infty$ in order to probe effects at large curvature. Here we will instead look 
for solutions involving dynamical higher order terms beyond the one-loop approximation.

\subsection{Quadratic solutions}
\label{sec:glqsol}

Exploiting the small curvature approximation, it is possible to look for quadratic solutions
where the $r^2$ term is dynamical.  Following \cite{opv}, we look for such solutions 
for values of $\alpha, \beta$ and $\gamma$ determined by requiring $\varphi(r)$ is quadratic. 

They are obtained by plugging into the FP equation for small $r$ approximation the ansatz
\bea
\varphi(r)=g_0+g_1 r+g_2 r^2,
\ena
and writing the equation as $\frac{N}{D}=0$.
Here $N$ is a polynomial of fifth order in $r$ and $N=0$ can be solved
for the six unknowns $\alpha, \beta, \gamma, g_0, g_1$ and $g_2$.
Since we are making a small $r$ approximation, we know that these solutions cannot in fact be global
but can give quite good approximation to the exact solutions.
Of course for the solution to be consistent with the exact equation,
the bounds~(\ref{zmb}) must be satisfied.

We then find the following distinct solutions listed in~Table~\ref{t1} where the critical
exponents have been found using polynomial expansions around $r=0$ up to order $n$ as indicted.
Only 6th and 7th solutions have stable critical exponents $\t$ less than or equal to 4
and are reliable. Both have three relevant critical exponents.
The third one appears to have critical exponents less than or equal to 4, but they are
not stable and not reliable.
\begin{table}[h]
\begin{center}
\scalebox{0.88}{
\begin{tabular}{|c|c|c|c|c|c|c|c|c|l|}
\hline
$\a$&$\b$&$\c$&$10^3 g_0$&$10^3 g_1$&$10^3 g_2$& $\t\;(n=7)$& $\t\;(n=8)$ & $\t\;(n=9)$ \\
\hline
$-0.441$ & $-0.0460$ & $-0.129$ & 9.42 & $-3.80$ & 0.721 & 0.389 & 3.24 & 433, 0.776 \\
$-0.463$ & $-0.0468$ & $-0.0468$ & 9.33 & $-4.62$ & 0.877 & 0.301 & 1.73 & 153, 0.783 \\
0.767 & 0.250 & 1.18 & 5.86 & $-2.59$ & 0.589 & 0.604 & $3.54, 0.706$ & 0.359 \\
1.85 & 3.09 & 2.27 & 3.42 & $8.97$ & 2.84 & 8.02 & 7.80 & 7.54 \\
0.805 & 0.308 & $-0.238$ & 5.40 & $5.23$ & $-1.28$ & 7.29 & $8.47 \pm 2.47i$ & 7.08 \\
$-0.497$&$-4.22$&$0.278$&$2.96$&$-16.6$&$-0.235$& 2.94, 0.980 & 2.94, 0.982 & 2.94, 0.984 \\
$-0.266$&$-17.8$&$0.252$&$2.91$&$-12.7$&$-0.0119$& 2.76, 1.75 & 2.76, 1.74 & 2.76, 1.74 \\
$-0.683$ & $-0.102$ & $-0.165$ & 6.92 & $-9.63$ & 2.00 & 12.0, 5.30 & $0.00326, 12.6\pm 17.5i$
 & 8.92   \\
$-1.13$ & $-0.432$ & $-0.354$ & 4.67 & $-17.8$ & 5.38 & $6.29\pm 1.48i$ & $6.07 \pm 0.68i$
 & 9.68, 4.12 \\
2.21& 3.24 & 1.17 & 3.48 & $18.7$ & 8.56 & $4.09$ & $4.05$ & 4.00 \\
\hline
\end{tabular}
}
\end{center}
\caption{Quadratic solutions of the spectral sum FP equation. In the last column,
we report the results for the positive (real part of) critical exponents, evaluated up to 9th order
polynomial expansion.
The critical exponent $4$ is present in all solutions and is related to the cosmological term.
Those solutions with critical exponents larger than 4 are not reliable.
}
\label{t1}
\end{table}

Unfortunately, the bounds Eq.~(\ref{zmb}) on $\a$ and $\b$ are violated for the sixth FP
as is the bound on $\b$ for the seventh FP. As a consequence these solutions do not make sense
globally even if we were to integrate towards larger curvature.

\subsection{Polynomial solutions}

We next study polynomial solutions to the flow equation which can be compared with those obtained
on compact spacetimes. To this end, we look for solutions of the form:
\bea
\vp_p(r) = \sum_{i=0}^n g_i r^i,
\ena
for fixed endomorphism parameters $\a, \b$ and $\c$.
In particular we take the case 
where the reference operator is $-\bnabla^2$ for all modes,
and a type-II cutoff where the reference operator contains
precisely the $\bR$-terms that are present in the Hessian.
In addition we also present solutions for the choice~\p{cut3}.
For each of these cases, we have looked for convergence of the couplings and critical exponents
as the order $n$ is increased. Unlike the case for the compact spacetime~\cite{fallslitim,dsz2,opv,opv2},
we do not find good convergence of the values of the couplings or critical exponents from order to order.
An interesting observation, though, is that the coefficients for higher order terms than quadratic
are in general very small.
For example, using the type I cutoff we have at order $n=8$ solutions
%
%
\bea
\vp_p(r) &=& 0.00346 - 0.00940 r -0.00371 r^2 + 5.96 \times 10^{-5} r^3 + 3.19 \times 10^{-4} r^4 \nn
&& +2.67 \times 10^{-5} r^5-1.71 \times 10^{-5} r^6 - 8.46 \times 10^{-8} r^7 +\cdots,
\label{solIa}
\ena
with the three relevant eigenvalues of the stability matrix $-41.5$, $-4$, $-1.69$
and
\bea
\vp_p(r) &=& 0.00389 - 0.00197 r + 0.00168 r^2 - 2.84 \times 10^{-4} r^3 - 8.44 \times 10^{-5} r^4 \nn
&& -7.80 \times 10^{-6} r^5- 1.15 \times 10^{-5} r^6 - 1.01 \times 10^{-6} r^7 +\cdots,
\label{solIb}
\ena
with just two relevant eigenvalues $-4$, $-0.825$.
Extending the order to $n=9$, we then find solutions
\bea
\vp_p(r) &=& 0.00397 - 0.00373 r +0.00241 r^2 - 8.74 \times 10^{-5} r^3 -1.46 \times 10^{-4} r^4 \nn
&& -1.43 \times 10^{-5} r^5-1.38 \times 10^{-5} r^6 - 1.05 \times 10^{-5} r^7 +\cdots, \nn
\vp_p(r) &=& 0.00331-0.00872 r-0.00157 r^2+0.00151 r^3+0.000591 r^4 \nn
&&+0.0000737
   r^5+0.000106 r^6+0.0000746 r^7+0.0000225 r^8+0.0000126 r^9, \nn
\vp_p(r) &=& 0.00254-0.00971 r+0.000508 r^2+0.000425 r^3-0.000833 r^4 \nn
&& + 0.00160 r^5-0.0116 r^6-0.0293 r^7-0.000213 r^8+0.0546 r^9.
\ena
 The relevant eigenvalues have real parts given by $-21.21$, $-21.21$, $-4$, $-1.46$
for the first FP, $-4.85377$, $-4$, $-2.02686$ for the second FP
and $-4,-3.57868,-3.57868,-3.42532,-2.50629$ for the third FP.
When there are two same eigenvalues, this means that they are a complex conjugate pair and
we show just the real parts.

Using a type II cutoff with  $\a=-\frac{1}{6}$, $\b=\frac{1}{3}$ and $\c=\frac{1}{4}$,
we find for $n=8$ the fixed points
\bea
\vp_p(r) &=& 0.00343 - 0.0120 r -0.00390 r^2 + 5.20 \times 10^{-5} r^3 + 2.93 \times 10^{-4} r^4 \nn
&& +2.71 \times 10^{-5} r^5-3.10 \times 10^{-5} r^6 - 3.46 \times 10^{-6} r^7 +\cdots,
\ena
with the three relevant eigenvalues $-13.3$, $-4$, $-1.65$, and
\bea
\vp_p(r) &=& 0.00398 - 0.00727 r + 0.00454 r^2 + 6.63 \times 10^{-4} r^3 - 1.66 \times 10^{-4} r^4 \nn
&& -2.75 \times 10^{-5} r^5 + 2.48 \times 10^{-5} r^6 - 2.63 \times 10^{-5} r^7 +\cdots,
\label{solIIb}
\ena
with the eigenvalues $-9.97$, $-4$, $-2.06$.
\\

Using the cutoff with~\p{cut3} $\a=-\frac{17}{48}$, $\b=-\frac{9}{48}$, $\c=-\frac{13}{48}$,
we find for $n=8$
\bea
\vp_p(r) &=& 0.003310 - 0.01206 r -0.002131 r^2 + 5.210\times 10^{-4} r^3 +5.697 \times 10^{-4} r^4 \nn
&& +2.109 \times 10^{-4} r^5+3.899 \times 10^{-5} r^6 + 1.700 \times 10^{-6} r^7
+5.435 \times 10^{-6} r^8,\nn
\vp_p(r) &=& 0.003273-0.01205 r-0.001767 r^2+8.960 \times 10^{-4} r^3+7.664 \times 10^{-4} r^4 \nn
&& +2.830 \times 10^{-4} r^5+8.320 \times 10^{-5} r^6+2.885 \times 10^{-6} r^7+2.814 \times 10^{-5} r^8, \nn
\vp_p(r) &=&  0.003237-0.01215 r-0.001483 r^2+0.001351 r^3+0.001503 r^4  \nn
&&+2.831 \times 10^{-4} r^5+8.360 \times 10^{-4} r^6-1.618 \times 10^{-4} r^7+ 6.824 \times 10^{-4} r^8\,,
\ena
with the negative eigenvalues for the stability matrix $-3.98$, $-4$ for the fist solution, and
$-4.09$, $-4$, $-1.05$ for the second and $-4$, $-3.05$, $-1.33$, $-1.33$  for the third.

For $n=9$, we find the fixed points
\bea
\vp_p(r) &=& 0.003341-0.01214 r-0.002510 r^2+2.181 \times 10^{-4}  r^3+0.0004578 r^4
 +1.983 \times 10^{-4} r^5 \nn
&& +3.688 \times 10^{-5} r^6-1.275 \times 10^{-6} r^7- 1.430 \times 10^{-6}r^8+2.486 \times 10^{-6}r^9, \nn
\vp_p(r) &=& 0.003413 - 0.01259 r -0.003815 r^2 - 7.380\times 10^{-4} r^3 +1.469 \times 10^{-4} r^4 \nn
&& +2.142 \times 10^{-4} r^5+1.004 \times 10^{-4} r^6 + 1.489 \times 10^{-5} r^7
-8.428\times 10^{-6} r^8 -9.574 \times 10^{-6} r^9, \nn
\vp_p(r) &=& 0.003196-0.01241 r-0.001220 r^2+0.001926 r^3+0.008472 r^4-8.934 \times 10^{-4} r^5 \nn
&& +0.01177 r^6+0.01908 r^7-0.01366 r^8+0.1067 r^9, \nn
\vp_p(r) &=& 0.003253-0.01210 r-0.001601 r^2+0.001169 r^3+9.672 \times 10^{-4} r^4+4.247 \times 10^{-4}
   r^5 \nn
&&   +1.662 \times 10^{-4} r^6+1.238 \times 10^{-4} r^7+7.082 \times 10^{-5} r^8+5.341 \times 10^{-5} r^9 ,
\ena
with the negative eigenvalues for the stability matrix $-4$, $-3.14$ for the first solution,
$-9.00$, $-4$, $-1.16$ for the second
$-4$, $-3.07$, $-0.33$, $-0.33$ for the third and
$-5.27$, $-4$, $-2.18$ for the fourth fixed point, respectively.

In all these cutoff  schemes, the number of relevant directions varies from two to five,
and the degree of convergence of the solutions for the hyperbolic case seem to be poorer compared with
the compact case when we increase the number of the polynomial terms, though some of them look
to have better convergence.
This makes a sharp contrast to the polynomial solutions for compact spaces studied
before~\cite{cpr1,ms,cpr2,fallslitim,Eichhorn,opv,opv2}. However we cannot rule out the possibility
that convergence will be found at orders beyond those considered here.

\section{Global solutions in four dimensions}
\label{sec:Global}
We now turn our attention to global solutions to the flow equation in four dimensions.

\subsection{Asymptotically free solutions}
First we will look for solutions to the flow equation where the $R^2$ coupling is allowed to diverge.
In this way we can consider asymptotically free fixed points without truncating the $f(R)$ action.

To begin with, we consider the fixed point equation with $\dot{\vp}(r) = 0$ and where
we allow the $r^2$ coupling to diverge. To find the solution, we write
\be \label{ansatz}
\vp(r) = - \frac{ r^2}{b} + \phi(r) ,
\ee
and insert this ansatz into the fixed point equation.
We are interested in solving the fixed point equation for $b \to 0$ to obtain a finite part $ \phi(r)$.
To this end, we expand the equation in $b$ and it then takes the simple form:
\be
\label{AfreeFP}
4 \phi(r)  -2 r  \phi'(r) =  \frac{ N_2(r) \Theta \left(r+\frac{1}{\alpha
   +\frac{17}{48}}\right)}{ \alpha  r+\frac{r}{6}+1}+\frac{N_0(r) \Theta
   \left(r+\frac{1}{\beta +\frac{3}{16}}\right)}{\beta 
   r+1}-\frac{N_1(r) \Theta \left(r+\frac{1}{\gamma
   +\frac{13}{48}}\right)}{( \gamma -\frac{1}{4}) r+1}  + \mathcal{O}(b).
\ee
where we include the theta functions to make explicit the vanishing of the traces at the critical
values of curvature \p{range}.
Since the RHS of \p{AfreeFP} is independent of $\phi(r)$, the equation can be integrated
straightforwardly.
We observe that there are poles depending on the endomorphism parameters which in principle can
lead to a break down of the convexity of the effective action.
To remove them, we put $\a = -1/6$, $\b = 0 $ and $\c = 1/4$ which set all the denominators to $1$.
The contribution of the spin one degree of freedom then vanishes at $r = -48/25$
whereas the spin zero and spin two contributions vanish at $r = -16/3$.
We can then solve this equation~\p{AfreeFP} in the small $r$ regime, obtaining
\be \label{smallrphi}
\phi(r) = \frac{1}{2}  \frac{371 r^2 \log (r^2)}{23040 \pi ^2}-\frac{5 r}{64 \pi
   ^2}+\frac{3}{128 \pi ^2} + a r^2 ,
\ee
where $a$ is an integration constant. We note that the coefficient of the $r^2 \log(r^2)$ can be shown
to be universal i.e. independent of the choice of the endomorphisms and the cutoff function.
For a different choice of the endomorphism parameters, we would obtain additional terms
$\sim r^2 \log((4 \gamma -1) r+4)$, $\sim r^2 \log(6 \alpha  r+r+6)$ and $\sim r^2 \log (\beta r+1)$
as well as modified coefficients for the constant and linear terms.
The solution \p{smallrphi} can be analytically continued to positive curvature provided
the curvature is smaller than the radius on convergence $|r| < r_C$ for the small curvature expansion.

In the small $r$ approximation, we can look for constant curvature solutions $r=r_0$  to
the equation of motion \p{EofM} for either sign which in the dimensionless form corresponds to 
$4 \phi(r_0)  -2 r_0  \phi'(r_0) = 0$. We note that the $r^2$ term in $\varphi$ does not alter
the equation of motion. We then find a pair of solutions to this equation of motion
at positive and negative curvatures:
\be
\label{r0}
r_+= \frac{6}{371} \left(\sqrt{33630}-150\right)  \approx  0.539917
  \,,\,\,\,\,\,\,\, r_-=\frac{6}{371} \left(-150-\sqrt{33630}\right)  \approx -5.39167.
\ee
To determine whether $r_+$ is a physical solution, we need to verify that $r_+ < r_C$ where $r_C$
is the value of $|r|$ for which the small $r$ approximation breaks down.
To check whether $r_-$ is physical, we look at the exact form of \p{AfreeFP}.

\begin{figure}[t]
\centering
\includegraphics[width=1\textwidth]{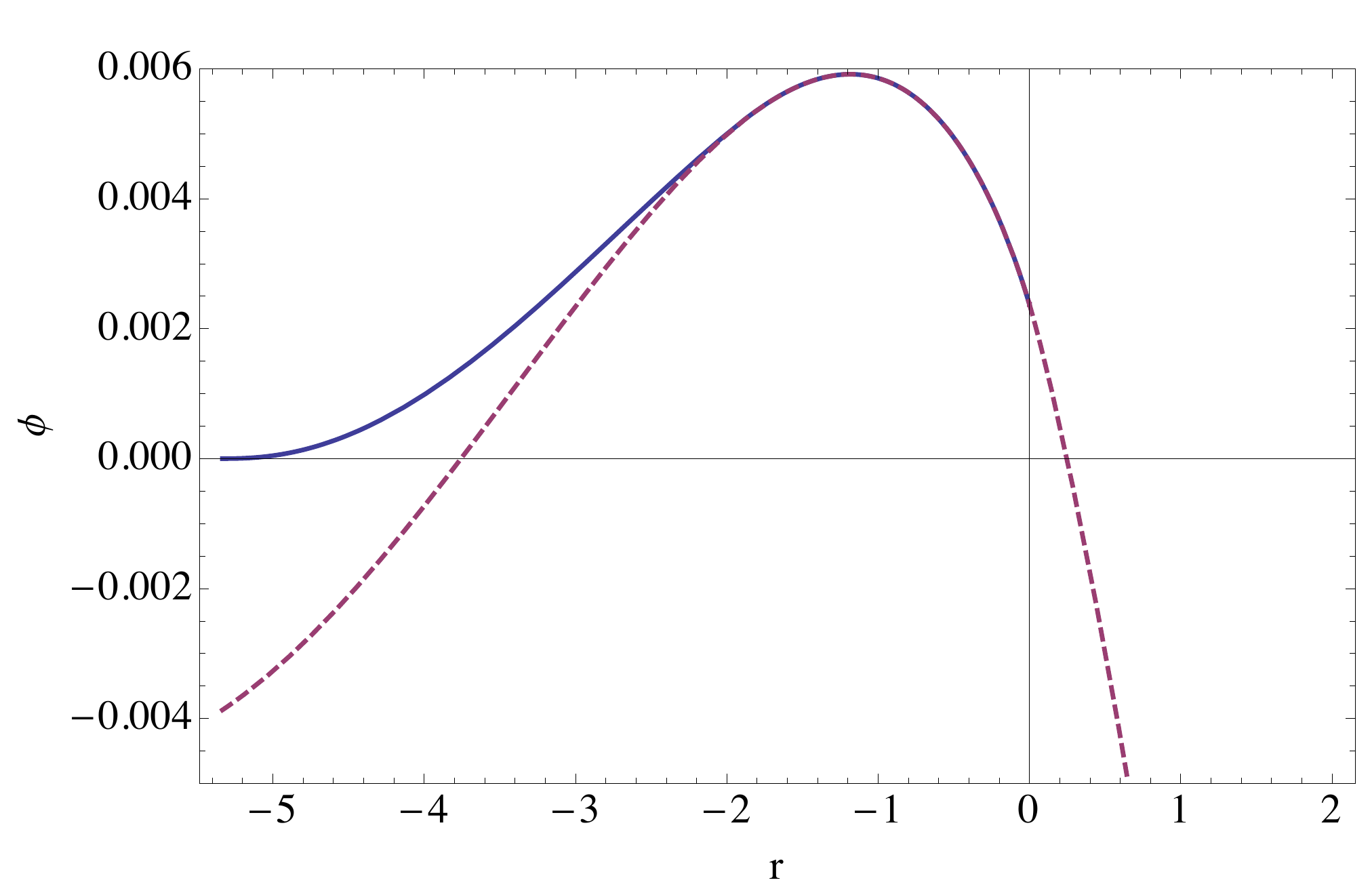}
\caption{Global solution for negative curvature with an asymptotically free $r^2$ coupling.
Boundary conditions are set so that $\phi(-16/3) = 0$, for the exact solution,
fixing the finite $r^2$ contribution.  The solid (blue) line is the exact solution whereas
the dashed (purple) line is the small curvature approximation which is continued for positive curvature.
We note that the vanishing of the solution for $r = -16/3$ is imposed by choosing
the constant of integration $a$ suitably.}
\label{phi}
\end{figure}

To study this, let us first obtain the global solution for $-\infty < r<0$ by integrating \p{AfreeFP}.
The plot of the obtained solution is shown in Fig.~\ref{phi} and compared to the small $r$ approximation.
We observe that the small curvature approximation is in agreement up till the point where
the first theta function vanishes at $r =  -48/25$. As such, an estimate for the radius of convergence 
of the small $r$ expansion is obtained as
\be
r_C \approx 48/25\,,
\ee
which indicates that the positive solution in \p{r0} is well within the radius of convergence.

\begin{figure}[t]
\centering
\includegraphics[width=1.\textwidth]{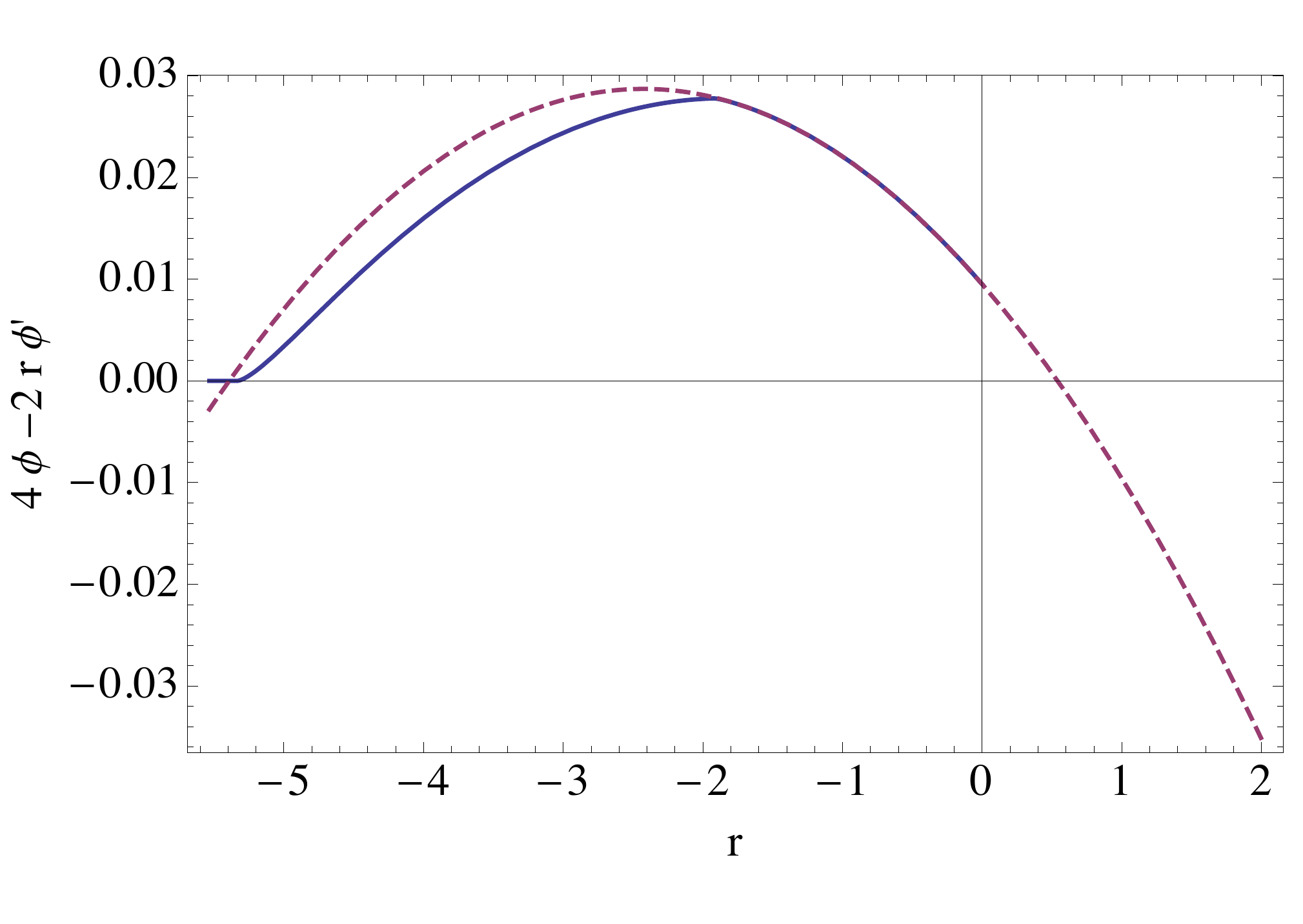}
\caption{The equation of motion for the asymptotically free fixed point. The solid (blue) line is
the exact result whereas the dashed (purple) line is the small curvature approximation.
We observe that up to the point where the first theta functions vanish, the small $r$ approximation
is in excellent agreement with the exact result. There are solutions for both positive and negative
curvatures. However the latter can be considered as an artifact of the optimised cutoff.  }
\label{EofM_phi}
\end{figure}

What about $r_-$? The absolute value $|r_-|$ is outside of this, and this is a first indication
that it may not be a physical solution.
Since the RHS of \p{AfreeFP} is zero at $r = - 16/3$, it follows that
the solution beyond this point is simply of $r^2$ form. Furthermore due to this vanishing,
we have a solution to the equation of motion at this point:
\be
\label{rminus}
r_-  = - 16/3 \approx -5.33333,
\ee
which is in good agreement with $r_-$ in the small curvature approximation~\p{r0}.
The value $4\phi-2r\phi'$ of the equation of motion is
plotted in Fig.~\ref{EofM_phi} which is found simply by plotting the RHS of \p{AfreeFP}.
The existence of this solution is due to the fact that the RHS of the flow equation vanishes
once the last modes is integrated out.
When at least one parameter takes its critical value \p{cut3}, the RHS never vanishes
for the whole finite range of negative $r$, and only the $r_- = - \infty$ can be a solution.
This means that $r_-$ is not generically finite.
We take this as a second evidence that it is not physical.

It is important to realize that the fixed point solution $\phi(r)$ with a free $r^2$ coupling
is not the most general solution to the flow equation found in the limit of $b \to 0$.
In fact we have implicitly assumed that
\be \label{betab0}
\partial_t \frac{1}{b} = 0,
\ee
to obtain $\eq{AfreeFP}$ which follows from $ \dot{\vp}=0$. In this sense the solution is
a true fixed point. However when we expand $\phi(r)$ for small $r$, we will obtain a non-analytic
expression due to the $r^2 \log (r^2)$  term in  \p{smallrphi}.
This suggests the existence of a different solution to the flow equation without a strict
scaling solution condition ($\dot{\vp} = 0$). Instead, we look for a solution with the beta function
for the $R^2$ couplings
\be
\label{betab}
\beta_b \equiv \partial_t b = - w b^2 + \mathcal{O}(b^3),
\ee
where we assume $w \neq 0$.
Such a fixed point obtained in the $f(R)$ approximation can be understood as an approximation to
the well-known asymptotically free fixed point in curvature squared
gravity~\cite{JT,ft,AB,Codello:2006in,Niedermaier}.
Here since we include only the $R^2$ term, we will only get an approximation to the full theory. 
A similar approximation has been considered in \cite{Copeland} where only the $R$ and $R^2$ terms were 
retained revealing an asymptotically free fixed point. 
Here we are not limited by the small curvature expansion and can therefore find
a more general solution of the $f(R)$ type. To this end, we write an ansatz for this solution as
\be
\vp(r) = -\frac{ r^2}{b} + \tilde{\phi}(r) ,
\ee
where for small $b$ we further demand that $\partial_t b = \beta_b$ be given by \eq{betab}.
Inserting this ansatz into the flow equation with $\partial_t \tilde{\phi}(r)= 0$ and again
taking the limit $b\to 0$, the flow equation then takes the form 
\be \label{AfreeFPw}
4 \tilde{\phi}(r)  -2 r  \tilde{\phi}'(r) - w r^2 =  \frac{ N_2(r) \Theta \left(r+\frac{1}{\alpha
 +\frac{17}{48}}\right)}{ \alpha  r+\frac{r}{6}+1}+\frac{N_0(r) \Theta
 \left(r+\frac{1}{\beta +\frac{3}{16}}\right)}{\beta r+1}-\frac{N_1(r) \Theta \left(r+\frac{1}{\gamma
+\frac{13}{48}}\right)}{( \gamma -\frac{1}{4}) r+1}  + \mathcal{O}(b) \,,
\ee
for small $b$. To determine $w$, we now demand that $\tilde{\phi}(r)$ be analytic in the limit $r\to 0$
which means that for small curvature, $\tilde{\phi}(r)$ has a polynomial expansion
\be
\label{smallrphi_expansion}
\tilde{\phi}(r) \sim \sum_n \tilde{g}_n r^n,
\ee
for finite coefficients $\tilde{g}_n$. Expanding \p{AfreeFPw} for small $r$, we then have
\be
4 g_ 0 + 2 g_1 r  - r^2 w + \cdots =  \frac{3}{32 \pi ^2}+\frac{(60 \a +12 \b -36 \c -41)}{384\pi^2} r
-\frac{371 }{11520 \pi ^2} r^2 +\cdots .
\ee
It follows that the coefficient of the universal $r^2$ term determines the value of the coefficient
appearing in \eq{betab} to be
\be
\label{w}
w = \frac{371}{11520 \pi ^2},
\ee
independently of $\alpha$, $\beta$ or $\gamma$. This differs from the coefficient $\frac{1117}{8640\pi^2}$
found in \cite{Copeland} where the equation was evaluated on a sphere.
We then observe that $b$ is asymptotically free coupling which behaves as
\be
\label{b}
\frac{1}{b} = \frac{371 \log (k^2/\mu^2)}{23040 \pi ^2}.
\ee
Setting $\a = -1/6$, $\b = 0 $ and $\c = 1/4$, we remove the other non-analytic terms as before.
Now we have the solution~\p{smallrphi_expansion} for small $r$ given by
\be
\tilde{\phi}(r) =\frac{3}{128 \pi ^2}  -\frac{5 r}{64 \pi^2}+ a r^2 ,
\label{sol327}
\ee
without the logarithmic term.
In fact the solution $\tilde{\phi}(r)$ is related to the the solution $\phi(r)$ by
\be
\tilde{\phi}(r) = \phi(r) - \frac{1}{2} \frac{371 r^2 \log (r^2)}{23040 \pi ^2},
\ee
for all $r$ since both \p{AfreeFP} and \p{AfreeFPw} are linear differential equations which
differ only by the term $- w r^2$. Here the additional term is expected to be universal
since it carries the coefficient of the logarithmic divergence.

\begin{figure}[t]
\centering
\includegraphics[width=1.\textwidth]{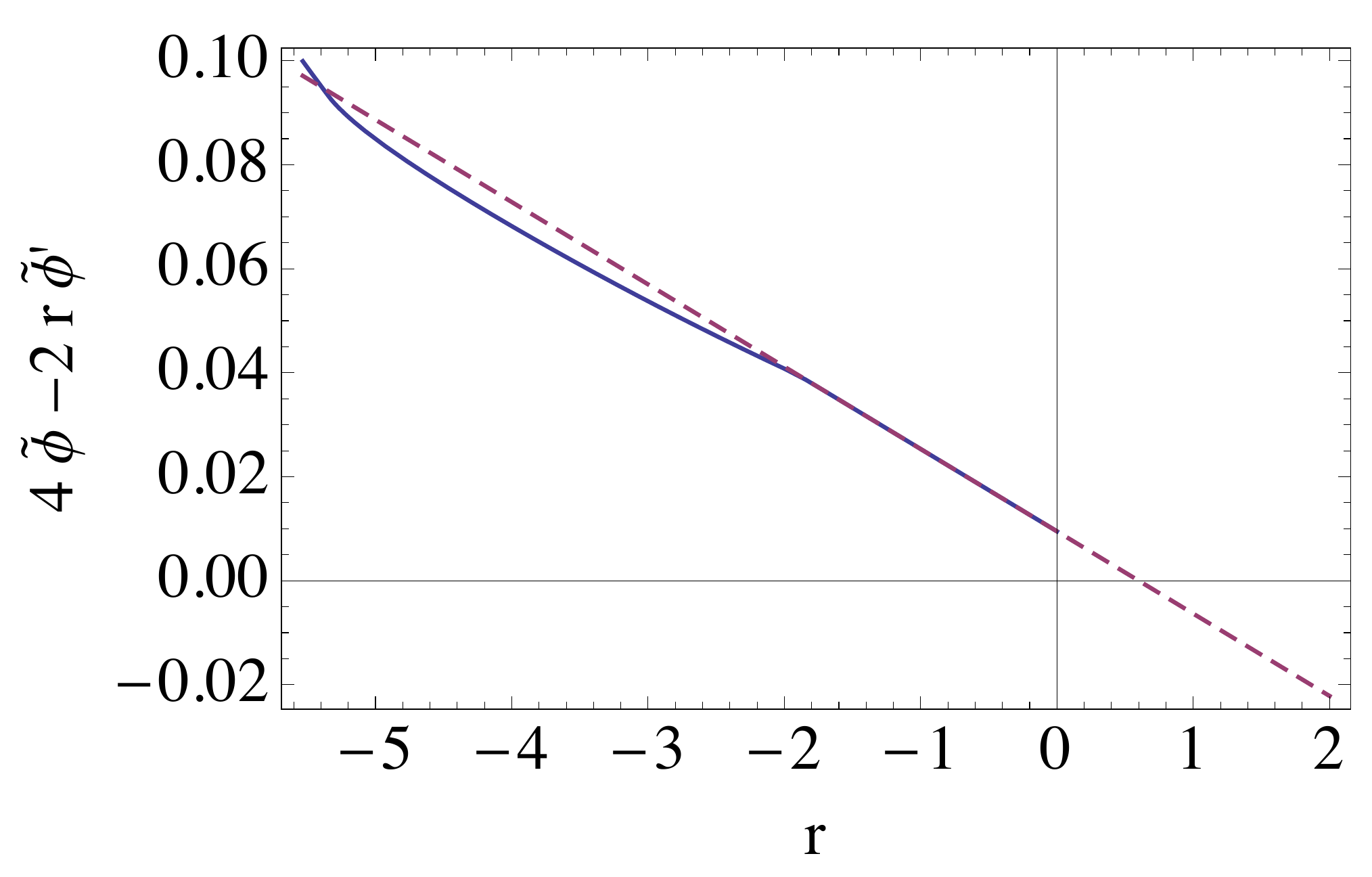}
\caption{The equation of motion for the asymptotically free fixed point approached with the $R^2$
coupling is governed by \p{betab} with the universal coefficient \p{w}. While there are no solutions
for negative curvature the analytical continuation of the small $r$ approximation (dashed line)
reveals a solution at positive curvature $r_0 = 3/5$.}
\label{EofM_tildephi}
\end{figure}

We can look again for constant
curvature solutions to the equation of motion \p{EofM} which are then solutions to
$4 \tilde{\phi}(r_0)  -2 r_0  \tilde{\phi}'(r_0) = 0$
which receive an extra term $w r^2$ compared to the equation of motion for $\phi$.
In the small $r$ approximation, using \p{sol327}, we have
\be
4 \tilde{\phi}(r)  -2 r  \tilde{\phi}'(r) \approx  \frac{3}{32 \pi ^2}-\frac{5 r}{32 \pi ^2},
\ee
which now only has a solution for positive curvature $r_0 = 3/5$.
In Fig.~\ref{EofM_tildephi} we plot the exact form of $4 \tilde{\phi}(r) -2 r  \tilde{\phi}'(r)$
observing that there is no solution for negative curvature.
To check the stability of the de-sitter solution at $r_0 = 3/5$, we can compute the
mass of the scalaron $m$ by evaluating $m^2 \equiv U''(\phi_0)$ given by \p{U''} at the minimum
corresponding to $R_0 =\frac{3}{5} k^2$. The stability of de-Sitter space requires $m^2 >0$.
Explicitly we find
\be
\frac{m^2}{k^2} \equiv  \frac{1}{k^2} \frac{1}{3 f''(R_0) } ( f'(R_0) - R_0 f''(R_0) )
=  \frac{5}{384 \pi^2}   \frac{23040 \pi ^2}{371 \log (k^2/\mu^2)},
\ee
where the subleading dependence on $a$ is neglected. This is positive in the UV limit as a consequence
of $w >0$. Additionally we have $f'(R_0)<0$
which is also needed for the graviton to have the right sign kinetic term.

Lowering the cutoff scale such that $R/k^2 < -16/3$, the RHS of the flow equation vanishes and
we then obtain the effective action: 
\be
\label{IR_fR}
\int d^4x \sqrt{g} f(R)= \int d^4x  \sqrt{g} \left( a R^2 -\frac{371 R^2 \log (R^2/\mu^4 )}{46080 \pi ^2}
\right),
\ee
where the $k$-dependence has cancelled between the $R^2/b$ term and the $R^2 \log(R^2/k^4)$ term.
Here we can take $a$ as a free parameter and fix $\mu$ to some fixed reference scale.
We conclude that in the UV, the solution is asymptotically free with a coefficient \p{w} and
that in the IR, the action stops running to take the form \p{IR_fR}.
Thus we have an effective action obtained after all quantum corrections have been integrated out.

Since the fixed point here is perturbative, it is straightforward to find the critical exponents.
Adding a scale-dependent perturbation to the solution $\tilde{\phi} + e^{t \theta} \delta \vp(r)$,
one finds that the perturbations are simply solutions to
\be
\theta \delta \vp(r)  + 4 \delta \vp(r) - 2 r  \delta \vp'(r) =0 ,
\ee
and thus the only analytic solutions are integer powers of $\delta \vp(r) \propto r^n$ with
the eigen-spectrum
\be
\theta = -4 + 2 n \,.
\ee
Hence in addition to the asymptotically free $R^2$ coupling, there are two further relevant couplings
corresponding to the vacuum energy and Newton's constant. The requirement of asymptotic safety then demands
that we set all irrelevant perturbations to zero, while the relevant perturbations generate
an Einstein-Hilbert action which we can add to \p{IR_fR} to obtain the effective action
\be
\int d^4x  \sqrt{g} f(R)   = \int d^4x \left( \frac{1}{16 \pi G} (-R + 2 \Lambda) + a R^2
-\frac{371 R^2 \log (R^2/M_{Pl}^4 )}{46080 \pi ^2}  \right)\,.
\ee
Here we have fixed the reference scale to the Planck mass $\mu = M_{Pl} \equiv 1/\sqrt{G}$
without loss of generality. We then have three free parameters $a$, $G$ and $\Lambda$.
This effective action (with $\Lambda = 0$ and $f \to -f$ after rotation to Lorentzian signature)
has been studied in \cite{Ben-Dayan:2014isa} in
the context of inflation and gives rise to a hilltop inflation.

\subsection{Global numerical solutions}
\label{sec:glsol}

We can also analyse the flow equation given in Eq.~\p{erge} at fixed endomorphisms,
and search for global numerical solutions with finite couplings.
For this purpose, it is convenient to choose the parameters~\p{cut3} such that the RHS of
the flow equation is nonzero over the whole range $-\infty <r <0$ without singularities.
The flow equation then takes the same form for the whole
range of $r<0$, and we can try to study the asymptotic expansion for large $r$ using \p{criticalexp}.
The expansion takes the form 
\bea
\vp(r) &=&  a r^2  + \sum_{n=1}^{\infty} a_i(a) \frac{1}{(-r)^{i/2}} \,,
\label{asy_series}
\ena
where $a$ is an arbitrary number.
Explicitly we find
\bea
\vp(r) &=& a r^2 +\frac{323}{1500\pi}\frac{\sqrt{3}}{(-r)^{1/2}}
-\frac{26648+14535\pi^2}{39375\pi} \frac{\sqrt{3}}{(-r)^{3/2}} \nn
&&
+\frac{8 (9181648 + 3975660 \pi^2 + 3270375 \pi^4)}{26578125 \pi} \frac{\sqrt{3}}{(-r)^{5/2}} +\cdots\,.
\label{asy}
\ena
At higher orders in the asymptotic expansion, one finds that the coefficients of $a_i$ become dependent on $a$
as indicated in \p{asy_series}.
The large negative curvature expansion therefore gives a one parameter family of solutions
in the asymptotic limit.

As a consequence of the asymptotic behaviour, global solutions are reduced to a one parameter family
despite the equation being third order. By numerically integrating the fixed point equation
starting at large $-r$ we can
then give initial conditions for different values of $a$ matching the asymptotic expansion.
Furthermore at $r=0$, we have to impose a regularity condition
which removes the remaining one free parameter since we then need to tune $a$.
For finite $r$, there are no further fixed singularities
which would otherwise over constrain the equation. As a result, we expect at most a discrete number of
regular global solutions, at least for the choice~\p{cut3}.

We have not been able to manage to find any global solutions of this type numerically.
However, a full numerical study is beyond the scope of this paper and we cannot rule out there existence. 
This is left for future study.

\section{Summary and discussions}
\label{sec:discussion}

In this paper we have derived a nonperturbative flow equation for quantum gravity on hyperbolic
spacetimes of constant curvature. The equation assumes that the scale-dependent effective action
is of the $f(R)$ form and hence is not sensitive to tensor structures which fall outside of
this approximation.
While such tensor structures are irrelevant on the LHS of the flow equation for maximally symmetric
spacetimes, curvature squared terms will alter the form of the flow equation 
since they contribute to $\Gamma^{(2)}_k$. 

Although the flow equations for $f(R)$ gravity have been studied previously,
these were studied on spheres where the traces, evaluated with the optimised cutoff, are discontinuous
due to the step functions. Consequently, additional approximations are made to smoothen the flow equation.
A key difference for negative curvature spacetimes is that the spectrum of the laplacian is continuous
and we therefore do not need to further approximate the traces when using the optimised cutoff.
In particular, we observe that once $k^2$ is less than the smallest eigenvalue the RHS of the flow equation vanishes
indicating the IR limit has been obtained. In this limit we have also noted that the equation becomes non-analytic.
One should note that the non-analytic form is the price which is paid for being able to evaluate
the traces explicitly with the optimised cutoff.       

For small curvature, we have found that the flow equation is well approximated by the equation
where the early-time heat kernel expansion is exploited.
Here we have found that the flow equations have a quite similar structure to the case of
compact spacetimes.
Importantly, however, it is clear that
this approximation is not applicable globally and breaks down for finite curvature.
Nonetheless this approximation is appropriate for studying polynomial solutions around $r=0$.
Here another difference from previous equations on the sphere arises since the early-time heat
kernel expansion is not modified by the removal of certain modes, as it is on the sphere.
It is important to point out that the differences come from the choice of a maximally symmetric
background and that it is therefore clear that they will not arise when a generic background is chosen.
While these modifications may seem minor, we find that they have a dramatic effect;
here we have found rather poor convergence of the polynomial solutions of increasing order.
Furthermore, we have observed that the requirement
that all modes are integrated out for $k \to 0$ puts a constraint on the form of the differential
operator entering the regulator. Therefore the quadratic small curvature solutions which do
show convergence (for the eigenvalues) are not permissible since the regulator will not vanish,
even in the limit $k \to 0$.

Here we have also studied the flow equation in the case for which the $R^2$ coupling is
asymptotically free where exact solutions exist. By requiring the analytical behaviour of
the solution for $R/k^2 \to 0$, we find that the beta function for the $R^2$ coupling is determined
by the universal coefficient. Then by taking $k \to 0$, the effective action is also determined
to contain an $R^2 \log(R^2)$ type correction. This fixed point can be understood as the projection
of the general curvature squared fixed point onto the $R^2$ coupling alone.
One therefore expects that on a more general backgrounds the effective action will take a more
involved form. It is nonetheless pleasing to see how quantum corrections to the effective action
are produced in approximations where the form is not truncated to a finite number of terms. 
It remains to see whether other fixed point solutions may be uncovered to the $f(R)$ flow equation
studied here. An analysis of free parameters suggests that there may be at most a discrete number
of solutions.

\section*{Acknowledgment}
We would like to thank Atsushi Higuchi for useful correspondence.
This work was supported in part by the Grants-in-Aid for
Scientific Research Fund of the Japan Society for the Promotion of Science (C) No. 24540290 and 16K05331, 
and the European Research Council grant ERC-AdG-290623.

\appendix

\section{Spectrum of Laplacian on hyperbolic space}
\label{spectrum}

The curvature tensors satisfy
\bea
\br_{\mu\nu\rho\s} = \frac{\br}{d(d-1)} (\bg_{\mu\rho}\bg_{\nu\s}-\bg_{\mu\s}\bg_{\nu\rho}),~~~
\br_{\mu\nu}=\frac{1}{d} \bg_{\mu\nu} \br,
\ena
with negative $\br=-\frac{d(d-1)}{a^2}$.
The eigenvalues of the Laplacian for spin $j$ on the hyperbolic space $H^d$ are continuous and
characterized by positive $\la$:
\bea
-\nabla^2 h^{(\la u)}_{\mu_1 \cdots\mu_j} = -\frac{\la^2+(\frac{d-1}{2})^2+j}{d(d-1)} \br
h^{(\la u)}_{\mu_1 \cdots\mu_j},
\ena
where $u$ is the discrete label for distinguishing eigentensors with the same eigenvalues~\cite{CH}.
See also \cite{Bene}.
We note that therefore an operator $-\nabla^2 - \alpha_j \br$ has a lowest eigenvalue 
\be \label{gap}
\delta_j = -\frac{(\frac{d-1}{2})^2+j}{d(d-1)} \br  - \alpha_j \br \,,
\ee
when $\lambda = 0$.
Eigentensors $h^{(\la u)}_{\mu_1 \cdots\mu_j}$ are normalized as
\bea
\int_{H^d} d^d x \sqrt{\bar g(x)} h^{(\la u)*} \cdot h^{(\la' u')}(x)=\d_{uu'} \d(\la-\la').
\ena
The analogue of the multiplicity for the continuous spectrum is the spectral function,
or Plancherel measure, defined by
\bea
\mu(\la) =\frac{\pi\Omega_{d-1}}{2^{d-2}g(j)} \sum_{u} h^{(\la u)*} \cdot h^{(\la u)}(x),
\ena
with  the volume of $S^{d-1}$
\bea
\Omega_{d-1} = \frac{2\pi^{d/2}}{\G(d/2)},
\ena
and the spin factor, the number of independent solutions, given by
\bea
g(j) =\frac{(2j+d-3)\cdot(j+d-4)!}{(d-3)!\, j!}.
\ena
[For $d=3$, $g(0)=1$ and $g(j)=2$ for $j\geq 1$.]
The spectral function is explicitly given by
\bea
\mu(\la) = \frac{\pi[\la^2+(j+\frac{d-3}{2})^2]}{2^{2(d-2)}\G(d/2)^2}\la\tanh(\pi\la)
{\prod_{j=1/2}^{(d-5)/2}(\la^2+j^2)},
\ena
for even $d\geq 4$ (for $d=4$ the product is omitted) and
\bea
\mu(\la) = \frac{\pi[\la^2+(j+\frac{d-3}{2})^2]}{2^{2(d-2)}\G(d/2)^2}
{\prod_{j=0}^{(d-5)/2}(\la^2+j^2)},
\ena
for odd $d\geq 3$.

The trace is given by integrating over the parameter $\la$ with the exponent $E_\la(d,j)$ of
the operator $\Delta$ weighted by their multiplicity~\cite{Bene}:
\bea
\mbox{Tr}_{(j)}[W(\Delta+E)] =   \int_{H^d} d^d x \sqrt{\bar{g}} \,
 \frac{2^{d-2}g(j)}{\pi\Omega_{d-1}}\Big(\frac{-\br}{d(d-1)}\Big)^{d/2}
\int_0^{\bar\la^{(j)}} d\la \,\mu(\la) W(\Delta_\la+E).
\ena
where we assume the integrand has support support for $0< \lambda < \bar{\lambda}^{(j)}$.

\section{Heat kernel expansions on constant curvature spacetimes}
\label{H4vsS4}

Let $W(-\nabla^2 - q R)$ be a function of the Laplace type operator such as the ones
appearing under the traces of Eq.~(\ref{frge}).
Here we wish to determine the trace over modes of a differentially constrained field
in terms of traces for unconstrained fields on constant curvature backgrounds.
Here we consider both the non-compact topology $H^d$ and the compact topology $S^d$.
We are interested in the following traces
\be \label{listoftraces}
\Tr_{0} '' \left[ W(-\nabla^2 - \beta R)\right]\,, \,\,\,\,\,  \Tr_{1T}' \left[
W(-\nabla^2 - \gamma R)\right] \,, \,\,\,\,\,\,\,\,\,\,  \Tr_{2T^2}  \left[ W(-\nabla^2
- \alpha R)\right] ,
\ee
where $\Tr_{0}''$ denotes the trace over scalar modes with the constant mode and
the CKVs removed, $\Tr_{1T}'$ denotes the trace over transverse-vectors
with the KVs removed and $\Tr_{2T^2}$ denotes the trace over symmetric
transverse-traceless tensors. In particular we are interested in the heat kernel coefficients
in four dimensions
\be
\Tr_{({\rm spin })} [e^{s (\nabla^2 + \beta R)}]
= \int d^4x \sqrt{g} \frac{1}{(4 \pi s)^2} ( b_0 + b_2 R s + b_4 R^2 s^2 + b_6  R^3 s^3 + ... ),
\ee
where $\Tr_{({\rm spin })}$ denotes one of the traces above. The coefficients $b_{n}$ are
listed in table~\ref{HKC} where we set $\zeta =1$ on $S^4$ and $\zeta = 0$ on $H^4$.
Differences occur because normalisable constant modes, KVs and CKVs exist
on the compact topology $S^d$ but do not occur on $H^d$.
Below we account for these differences denoting the number of these modes on the $S^d$ ($H^d$)
by  $\Sigma_{\rm const.}=1$ ($\Sigma_{\rm const.}=0$) for the constant modes,
$\Sigma_{KV} = \frac{1}{2} d (d+1)$ ($\Sigma_{KV} = 0$) for the  KVs
and $\Sigma_{CKV} = d+1$ ($\Sigma_{CKV} = 0$) for the CKVs.

\subsection{Scalars}
\label{H4scalar}

For the scalar trace in \eq{listoftraces} the difference between the heat kernel coefficients
for the two topologies is due to the removal of the constant mode and CKVs from the trace
\be
\Tr_{0} '' \left[ W(-\nabla^2 - \beta R) \right] = \Tr_{0}  \left[ W(-\nabla^2 - \beta R)]\right]
- \Sigma_{\rm const.}W(- \beta R)  - \Sigma_{CKV}  W\left(- \beta R + \frac{R}{d-1}\right) \,.
\ee
The corresponding heat kernel coefficients are given in table~\ref{HKC} on $S^4$ ($\zeta = 1$)
and $H^4$ ($\zeta = 0$).

\subsection{Vectors}
\label{H4vector}

First consider a vector $V_\mu$ and decompose it as
\be
V_\mu = V_\mu^T + \nabla_\mu \phi,
\ee
where $V_\mu^T$ is a transverse vector. The compete set of eigenmodes of $V_\mu$ is then
eigenmodes of $V_\mu^T$ and $V_\mu^L = \nabla_\mu \phi $.
From the equation
\be
- \nabla^2 \nabla _\mu \phi =  \nabla_\mu  \left(- \nabla^2  -  \frac{R}{d}  \right) \phi,
\ee
we can determine the spectrum of the longitudinal modes from that of a scalar field.
However we need to take into account that the constant mode of $\phi$ will not contribute to $V_\mu$.
It follows that we can express a trace over transverse vectors modes as
\bea \label{Tr1T}
{\rm Tr}_{1 T}' W(-\nabla^2 - \c R)
&=& {\rm Tr}_{1} W(-\nabla^2- \c R) - {\rm Tr}_{0} W\left(-\nabla^2 - \frac{R}{d} - \c R\right) \nn
 && +\Sigma_{\rm const.} W\left(- \frac{R}{d} - \c R\right) - \Sigma_{KV} W\left(\frac{R}{d}
- \c R\right) ,
\ena
where ${\rm Tr}_{1}$ is the trace over unconstrained vectors.

\begin{table}[h]
\begin{center}
\scalebox{0.9}{
\begin{tabular}{|c|c|c|c|c|c|}
\hline
Spin &$b_0$&$b_2$&$b_4$&$b_6$  \\
\hline
$0$ &$1$&$\beta +\frac{1}{6}$&$\frac{1080 \beta ^2+360 \beta -540 \zeta +29}{2160}$
&$\frac{45360 \beta^3+22680 \beta ^2-68040 \beta  \zeta +3654 \beta +18900 \zeta+185}{272160}$\\ \hline
$1T$ &$3$&$\frac{1}{4} (12 \gamma +1)$&$\frac{2160 \gamma ^2+360 \gamma -540 \zeta -67}{1440}$
&$\frac{181440 \gamma ^3+45360 \gamma ^2-136080 \gamma  \zeta -16884 \gamma +41580
   \zeta -4321}{362880}$ \\ \hline
$2T^2$ &$5$&$\frac{5}{6} (6 \alpha -1)$&$\frac{1}{432} \left(1080 \alpha ^2-360 \alpha
 +270 \zeta -271\right)$&$\frac{45360 \alpha ^3-22680 \alpha ^2+34020 \alpha  \zeta -34146 \alpha
 +7560  \zeta -7249}{54432}$ \\ \hline
\end{tabular}
}
\end{center}
\caption{ Heat kernel coefficients on $H^4$ with $\zeta=0$ and $S^4$ with $\zeta =1$
}
\label{HKC}
\end{table}

\subsection{Tensors}
\label{H4tensor}

Similarly for a tensor, we consider the decomposition
\be
h_{\mu\nu} = h^T_{\mu\nu} +  g_{\mu\nu} \phi +  \nabla_{(\mu} V_{\nu)}
- \frac{1}{d} g_{\mu\nu} \nabla^\rho V_\rho,
\ee
where $h^T_{\mu\nu}$ is transverse-traceless and the brackets denote the symmetrisation.
Then we have the following identity on a constant curvature spacetime:
\be
- \nabla^2 \left( \nabla_{(\mu} V_{\nu)} - \frac{1}{d} g_{\mu\nu} \nabla^\rho V_\rho \right)
=   \nabla_{(\mu} \left(- \nabla^2 -\frac{(d+1) R}{(d-1) d}\right) V_{\nu)}
- \frac{1}{d} g_{\mu\nu} \nabla^\rho  \left(- \nabla^2 -\frac{(d+1) R}{(d-1) d}\right) V_\rho ,
\ee
which allows us to express the trace over transverse-traceless modes in terms of a trace of a vector,
scalar and tensor. In this case we must remember to remove the KVs and CKVs from
the spectrum of $V^\mu$.
We therefore have the following relation between traces
\bea
{\rm Tr}_{2 T^2} W(- \nabla^2 - \a R)
&=&  { \rm Tr}_{2} W(- \nabla^2- \a R) - { \rm Tr}_{1} W\Big(- \nabla^2 - \frac{(d+1) R}{(d-1) d}
- \a R \Big)  - {\rm Tr}_{0}  W(- \nabla^2 -  \a R ) \nn
&+&  \Sigma_{KV} W\Big( - \frac{2R}{d(d-1)} -  \a R \Big)
+ \Sigma_{\rm CKV} W\Big( - \frac{R}{d-1} -  \a R \Big) ,
\ena
where ${ \rm Tr}_{2}$ is the trace over symmetric tensor modes.


\section{Spectral sum approach for $d=3$}
\label{sec:flowspect3}

It may be interesting to study the flow equation for odd dimension since the integrals are simpler.
For $H^3$, the trace is given by
\be
{\rm Tr}_{(j)} W(\Delta+E)=
  \int_{H^3} d^3 x \sqrt{\bar{g}} \, \frac{g(j)}{2\pi^2} \Big(\frac{-\br}{6}\Big)^{3/2}
\int_0^{\bar\la^{(j)}} d\la \frac{\pi(\la^2+j^2)}{4 \G(3/2)^2} W(\Delta_\lambda+E)\ ,
\label{spectral_sum3}
\ee
where the eigenvalues and the corresponding multiplicities given in Appendix~\ref{spectrum} are used.
The support of $R_k(\Delta_\lambda(d,j)+E)$ is restricted to the modes:
\bea
\bar{\la}^{(2)}\!=\sqrt{\frac{6}{-r}\!-3-6\a} \,, \quad
\bar{\la}^{(1)}\!=\sqrt{\frac{6}{-r}\!-2-6\c} \,, \quad
\bar{\la}^{(0)}\!=\sqrt{\frac{6}{-r}\!-1-6\b} \,, \quad
\label{upperlimits3}
\ena
The integrals extend up to the these upper bounds.

Using Eq.~\p{spectral_sum3} and \p{diml}, we obtain the following flow equation
\bea
&& \hs{-10}
\dot \vp -2r \vp'+ 3 \vp \nn
&=& \frac{c_1 (\dot\vp'-2r \vp'')+c_2 \vp'}{\vp'[3+(3\a+1)r ]}
+ \frac{c_3 (\dot\vp''-2r \vp''')+c_4 \vp''}{2[2+(2\b-1)r]\vp''+\vp'}
- \frac{c_5}{3+(3\c-1) r},
\label{erge3}
\ena
where the coefficients are simply given by
\bea
c_1 &=& \frac{(-r)^{3/2}}{24\sqrt{6}\pi^2} \Big[ \frac{r}{5} (\bar\la^{(2)})^5
+ 2\Big(1+\Big(\a+\frac{7}{6}\Big)r\Big) (\bar\la^{(2)})^3
+ 24 \Big(1+\Big(\a+\frac{1}{2}\Big)r\Big) \bar\la^{(2)} \Big], \nn
c_2 &=& \frac{(-r)^{3/2}}{24\sqrt{6}\pi^2} \Big[ \frac{r}{5} (\bar\la^{(2)})^5
+ 2 \Big(3+\Big( \a+\frac{7}{6}\Big)r\Big) (\bar\la^{(2)})^3
+ 24 \Big(3+\Big(\a+\frac{1}{2}\Big)r\Big) \bar\la^{(2)} \Big], \nn
c_3 &=& \frac{(-r)^{3/2}}{36\sqrt{6} \pi^2} \Big[ \frac{r}{5} (\bar\la^{(0)})^5
+ 2 \Big(1+\Big(\b+\frac{1}{6}\Big)r\Big) (\bar\la^{(0)})^3 \Big], \nn
c_4 &=& -\frac{(-r)^{3/2}}{36\sqrt{6} \pi^2} \Big[ \frac{r}{5} (\bar\la^{(0)})^5
- 2 \Big(1-\Big(\b+\frac{1}{6}\Big)r\Big) (\bar\la^{(0)})^3 \Big], \nn
c_5 &=& \frac{(-r)^{3/2}}{6\sqrt{6} \pi^2} \Big[ (\bar\la^{(1)})^3+3\bar\la^{(1)} \Big].
\label{coe_spectral3}
\ena
This result is exact and does not involve complicated functions, so that we can easily
study the properties of the flow equation. The equation can be compared to the flow equation
derived in \cite{Demmel:2014sga,dsz1} on a $H^3$ and $S^3$ in the conformally reduced approximation.


\end{document}